\pdfoutput=1  

 \documentclass[preprint]{aastex}

\shorttitle{Infall to Keplerian Disk in L1551 IRS 5}
\shortauthors{Chou et al.}

\begin{document}


\title{Transition from the Infalling Envelope to the Keplerian Disk \\around L1551 IRS 5}
\author{Ti-Lin Chou\altaffilmark{1,2}, Shigehisa Takakuwa\altaffilmark{2},
Hsi-Wei Yen\altaffilmark{2}, Nagayoshi Ohashi\altaffilmark{2,3}, \& Paul T. P. Ho\altaffilmark{2,4}}
\altaffiltext{1}{Department of Physics, National Taiwan University, No. 1, Sec. 4, Roosevelt Rd., Taipei 106, Taiwan; tlchou@asiaa.sinica.edu.tw}
\altaffiltext{2}{Academia Sinica Institute of Astronomy and Astrophysics, P. O. Box 23-141, Taipei 10617, Taiwan}
\altaffiltext{3}{Subaru Telescope, National Astronomical Observatory of Japan,
650 North A'ohoku Place, Hilo, HI 96720, USA}
\altaffiltext{4}{Harvard-Smithsonian Center for Astrophysics, 60 Garden Street, Cambridge, MA 02138, USA}

\begin{abstract}
We present combined SubMillimeter Array (SMA) + Atacama Submillimeter Telescope Experiment (ASTE) images of the Class I protobinary L1551 IRS 5 in the CS ($J$ = 7--6) line,
the submillimeter images of L1551 IRS 5 with the most complete
spatial sampling ever achieved ($0\farcs9$ -- $36\arcsec$).
The SMA image of L1551 IRS 5 in the 343 GHz dust-continuum emission is also presented,
which shows an elongated feature along the northwest
to southeast direction ($\sim$160 AU $\times$ 80 AU), perpendicular to the associated radio jets.
The combined SMA+ASTE images show that
the high-velocity ($\gtrsim$1.5 km s$^{-1}$) CS emission traces the structure of the dust component
and shows a velocity gradient along the major axis,
which is reproduced by a geometrically-thin Keplerian-disk model with a central stellar mass of $\sim$0.5 $M_{\odot}$.
The low-velocity ($\lesssim$1.3 km s$^{-1}$) CS emission shows an extended ($\sim$1000 AU)
feature that exhibits slight south (blueshifted) to north (redshifted) emission offsets, which is modeled with a rotating
and infalling envelope with a conserved angular momentum.
The rotational motion of the envelope connects smoothly to the inner Keplerian rotation at a radius of $\sim$64 AU.
The infalling velocity of the
envelope is $\sim$three times lower than the free-fall velocity toward the central stellar mass of 0.5 $M_{\odot}$.
These results demonstrate transition from the infalling envelope
to the Keplerian disk, consistent with the latest theoretical studies of disk formation.
We suggest that sizable ($r\sim$50--200 AU) Keplerian disks are
already formed when the protostars are still deeply embedded in the envelopes.
\end{abstract}

\section{Introduction}

\begin{sloppypar}
Keplerian disks are ubiquitous around T-Tauri (Class II) sources ($e.g.$, Simon et al. 2000),
and are considered
to be the nurseries of planetary systems. Around younger protostellar (Class 0--I) sources
there are $\sim$2000--10000 AU scale protostellar envelopes that often show rotating
and infalling gas motions \cite{oha96,oha97,mom98,tak03,tak07}.
Recent high-resolution
interferometric observations in (sub)millimeter molecular lines have found
Keplerian disks around protostellar sources
\cite{bri07,lom08,jor09,tak12,tob12,har13,mur13,tak13,yen13,yen14,har14,lee14,tak14,oha14}.
The masses ($\sim$0.01--0.03 $M_{\odot}$) and radii ($\sim$80--300 AU) of these Keplerian disks
around protostars
are comparable to those of circumstellar disks around T-Tauri stars \cite{gui98,gui99,and05,and07,sch09}.
These results show that formation of Keplerian disks proceeds in the innermost parts of the infalling envelopes.
\end{sloppypar}

It remains controversial as to how Keplerian disks form
out of the infalling envelopes around protostars. Mellon \& Li (2008; 2009), and Li et al. (2011) have performed
magnetohydrodynamic (MHD) simulations of self-similar, collapsing envelopes to investigate the formation
of Keplerian disks. They found that, for typical magnetic field strengths in interstellar clouds,
magnetic braking is so efficient that rotationally-supported disks cannot form.
On the other hand, recent theoretical simulations
by Machida \& Hosokawa (2013) and Machida et al. (2014) successfully produced rotationally supported disks,
with their masses comparable to the observed results.
Their argument is that the Ohmic dissipation and the dispersal of the surrounding infalling envelope
that anchors the magnetic field significantly reduce the effect of the magnetic field on the disk formation. 

To put stringent observational constraints
on the theories of disk formation, direct imaging of the transition from infalling envelopes
to Keplerian disks, and studies of evolutions of Keplerian disks
embedded in infalling envelopes, are crucial.
Takakuwa et al. (2013) have found infalling gas outside of the Keplerian disk around L1551 NE,
a Class I protobinary located $\sim$2$\farcm$5 north-east of L1551 IRS 5,
and revealed that the angular
momentum of the infalling gas is much smaller than that of the Keplerian disk.
Furthermore, the infalling velocity onto the central Keplerian disk is much smaller
than the free-fall velocity toward the central stellar mass of 0.8 $M_\odot$ as derived
from the Keplerian rotation. These results suggest a reduced infalling velocity
and possible magnetic braking which decreases the rotational angular momentum of the infalling material.
A slower infalling velocity than the free-fall velocity has also been identified in the envelope surrounding a Keplerian disk
around a Class 0/I protostar L1527 IRS \cite{oha14}.
On the other hand, Yen et al. (2014) found parabolic free-falling flows
where the rotational angular momentum is conserved enroute to the central Keplerian disk around L1489 IRS.
These observational reports are so far the only examples of the transition from the infalling envelopes to the Keplerian
disks. Hence, a general picture of the transition, and the evolutionary sequence of Keplerian disks
embedded in infalling envelopes, are not yet established.

In order to further study the formation mechanism of Keplerian disks,
we have performed SubMillimeter Array (SMA)\footnote{The SMA
is a joint project between the Smithsonian Astrophysical Observatory and
the Academia Sinica Institute of Astronomy and Astrophysics and is funded by the Smithsonian
Institution and the Academia Sinica.} observations of L1551 IRS 5,
one of the best-studied Class I protostars, in the submillimeter continuum and CS ($J$=7--6) emission.
The SMA CS (7--6) data were combined with the single-dish mapping data of CS (7--6) from the Atacama Submillimeter Telescope Experiment (ASTE)\footnote{The ASTE
project is driven by Nobeyama Radio Observatory (NRO), a branch
of National Astronomical Observatory of Japan (NAOJ), in collaboration
with University of Chile, and Japanese institutes including University
of Tokyo, Nagoya University, Osaka Prefecture University, Ibaraki
University, and Hokkaido University.} \cite{tak11}.
L1551 IRS 5 is a protostellar binary system with a projected separation of $\sim$0.3$\arcsec$
\cite{loo97,rod98,lim06},
located in Taurus at a distance of 140 pc \cite{ken94}.
The two binary sources, located to the north and south,
both drive active radio jets along the northeast to southwest direction \cite{rod03}.
Previous interferometric observations of L1551 IRS 5 in millimeter molecular lines have found
a $\sim$2500 AU-scale rotating and infalling envelope surrounding the protobinary
\cite{oha96,sai96,mom98}.
The rotational motion in the envelope shows a radial dependence
of $v_{rot} \propto r^{-1}$ at a radii of $\sim$300 -- 1000 AU,
which is consistent with the conserved angular momentum.
The infall velocity over those radii is also consistent with free-fall
toward a central stellar mass of 0.1 $M_{\odot}$.
For radii at $<$300 AU,
higher-velocity ($\gtrsim 0.7$ km s$^{-1}$) features are present,
which indicates that the central mass should be as high as 0.5 $M_{\odot}$ \cite{mom98}.
Our previous SMA observations of L1551 IRS 5 in CS (7--6) found rotation-dominant gas motion in the inner
$\sim$400 AU scale of the protostellar envelope, suggesting the presence of a Keplerian disk \cite{tak04}.
However, the lower angular resolution in the SMA early science study ($\sim$3\farcs2$\times$2\farcs0)
prevented us from unambiguously identifying the Keplerian disk.
Our new SMA observations at a subarcsecond resolution, when combined
with the single-dish ASTE data, enable us to investigate the spatial
and kinematical structures with the most complete spatial sampling
ever achieved ($\sim$0$\farcs$9 -- 36$\arcsec$).
With these data, we have successfully identified the inner Keplerian disk and the transition from the outer infalling envelope.


In the following, we describe our SMA observations (Section 2), our continuum and molecular-line results
(Section 3), and the analysis to identify the Keplerian disk and the infalling envelope (Section 4).
In the last section (Section 5), we will compare the properties of the Keplerian disk and envelope
around L1551 IRS 5 to those of the theoretical predictions and around the other protostars.
Then we discuss the formation and the evolutionary mechanism of Keplerian disks embedded in
infalling envelopes.

\section{SMA Observations and Combining with the ASTE Data}

We observed L1551 IRS 5 with the SMA in its subcompact configuration
on 2009 September 21, and in its extended configuration on 2012 December 23
and 2013 January 4. Details of the SMA are described by Ho et al. (2004).
In the observation with the subcompact configuration,
CS ($J$=7--6; 342.882857 GHz),
HC$^{15}$N ($J$=4--3; 344.200109 GHz), SO ($J_N$=8$_{8}$--7$_{7}$; 344.310612 GHz),
and HCN ($J$=4--3; 354.505473 GHz) lines, and
continuum emission at both the upper and lower sidebands
with the LO frequency of 348.754 GHz and the IF range of 4--6 GHz
were measured simultaneously.
In the observations with the extended configuration,
CS ($J$=7--6), C$^{18}$O ($J$=3--2; 329.3305453 GHz),
$^{13}$CO ($J$=3--2; 330.587965 GHz), and SO ($J_N$=8$_{8}$--7$_{7}$) lines,
and continuum emission at both the upper and lower sidebands
with the LO frequency of 336.902 GHz and the IF range of 4--8 GHz
were measured simultaneously.
The SMA correlator consists of a bank of the spectral windows (``chunks'')
with a bandwidth of 82 MHz, and there are totally 22 and 42 chunks
at each sideband in the observations with the subcompact and extended configurations,
respectively.
The chunk assigned to the CS line had a spectral resolution of 203.125 kHz,
corresponding to a velocity resolution of 0.178 km s$^{-1}$.
The raw visibility data were calibrated and flagged with MIR, which is an IDL-based data reduction
package \citep{sco93}.
All the chunks at both sidebands except for those assigned for the strong molecular lines
were combined to make a single continuum channel, providing the total continuum bandwidth
3.444 GHz in the subcompact observation and 6.560 GHz in the extended observations.
Both the continuum channels taken in the subcompact and extended observations were co-added,
and the continuum image was generated with uniform weighting of the measured visibilities.
In the present paper, we will focus on the results of
the continuum emission (hereafter 343 GHz continuum emission)
and the CS (7--6) line. The CS line was observed with the SMA subcompact and
extended configurations, and with ASTE (see below).
The SMA $^{13}$CO, C$^{18}$O, and SO data exhibit an emission component
centered on the protobinary, with a northwest (redshifted) to southeast (blueshifted)
velocity gradient, similar to that of the CS data. The velocity gradient in the HCN line
is not clear because of the hyperfine blending, and the HC$^{15}$N line is too weak to
identify the velocity structure.

\begin{figure}
\plotone{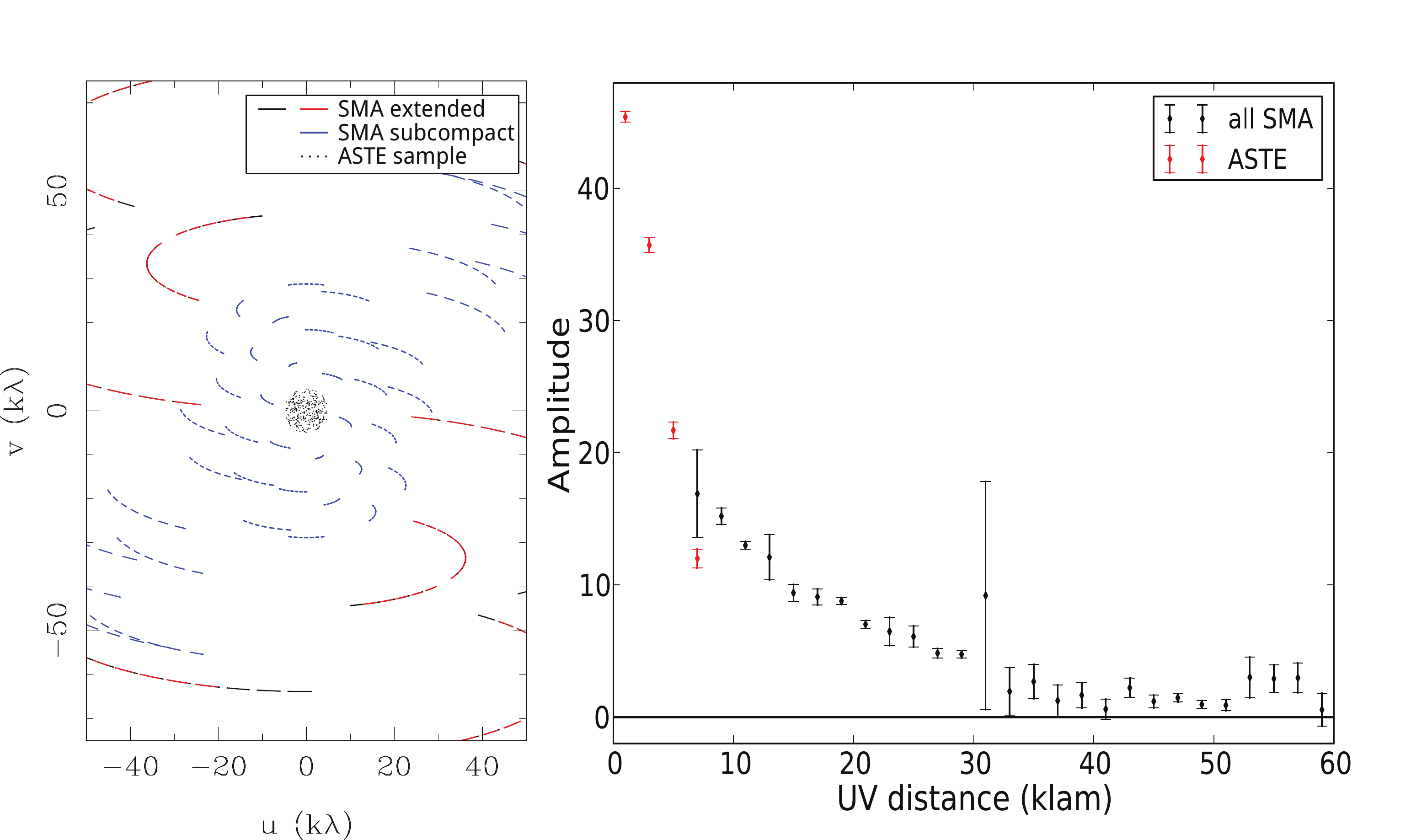}
\caption{Plots of the $uv$ coverages (\emph{left}) and the $uv$-distances v.s. amplitudes
of the SMA and ASTE datasets in the CS (7--6) emission (\emph{right}).
In the left panel, black and red lines represent the $uv$ coverages taken with the
SMA extended configuration at the two epochs, blue lines with the SMA subcompact configuration,
and black dots the $uv$ sample points adopted for the ASTE data. In the right panel,
the black data points show the mean values and the standard deviations of the SMA visibility data
and the red points those of the ASTE data at a 2 $k\lambda$ bin in the velocity range
from $V_{LSR}$ = 5.6 km s$^{-1}$ to 7.7 km s$^{-1}$.
\label{uv}}
\end{figure}

\begin{sloppypar}
We combined the SMA CS (7--6) data with the ASTE single-dish CS (7--6) data \citep{tak11},
adopting the method described by Takakuwa et al. (2007).
Details of the ASTE observations and results are described by Takakuwa \& Kamazaki (2011).
Since the diameter of the ASTE telescope (= 10 m) is larger than the minimum
projected baseline length of the SMA in its subcompact configuration ($\sim$6.7 m),
we were able to continuously sample the spatial scales from zero to $\sim$200 $k\lambda$.
To assess whether the combining these two datasets introduces any systematic uncertainty,
in Figure \ref{uv} we show the $uv$-distance and amplitude plot of the SMA and ASTE visibility
datasets. The visibility amplitudes of the inner ASTE data smoothly connect with the outer SMA
amplitudes, within the uncertainties of the absolute flux calibrations of $\sim$20\% in both datasets,
and thus combining the SMA and ASTE data does not introduce any additional uncertainty.
To obtain the most optimum combined image cube of the envelope and the disk in L1551 IRS 5,
we varied the number of the single-dish visibility sampling points and thus
the relative ratio of the density of visibility points per unit $uv$ area
between the single-dish and the interferometric data.
The ratio of 5 has been proposed as the most optimum ratio
of the single-dish to interferometric visibility densities by Kurono et al. (2009)
from their imaging simulations.
In our images, we found that the density ratio of $\sim$11 provides the best
compromise between the spatial resolution and sensitivity
and thus the most optimum combined image cube of the envelope and disk.
The combined image cube recovers $\sim$99$\%$ of the total CS (7--6) flux density
measured with ASTE at the spectral peak toward the field center
while the SMA only image contains $\sim$39$\%$,
and the combined image cube
can trace structures ranging from $\lesssim$0$\farcs$9 (combined synthesized beam)
up to $\sim$36$\arcsec$ (SMA field of view) without a significant effect of the missing flux.
In the following we discuss
the structure and kinematics of the envelope and disk with the
combined image cube adopting the ratio of 11.
Table 1 summarizes the parameters of the present SMA observations and
the SMA+ASTE image cube.
\end{sloppypar}

\placetable{tbl-1}

\section{Results}
\subsection{343 GHz Continuum}

\begin{figure}
\plotone{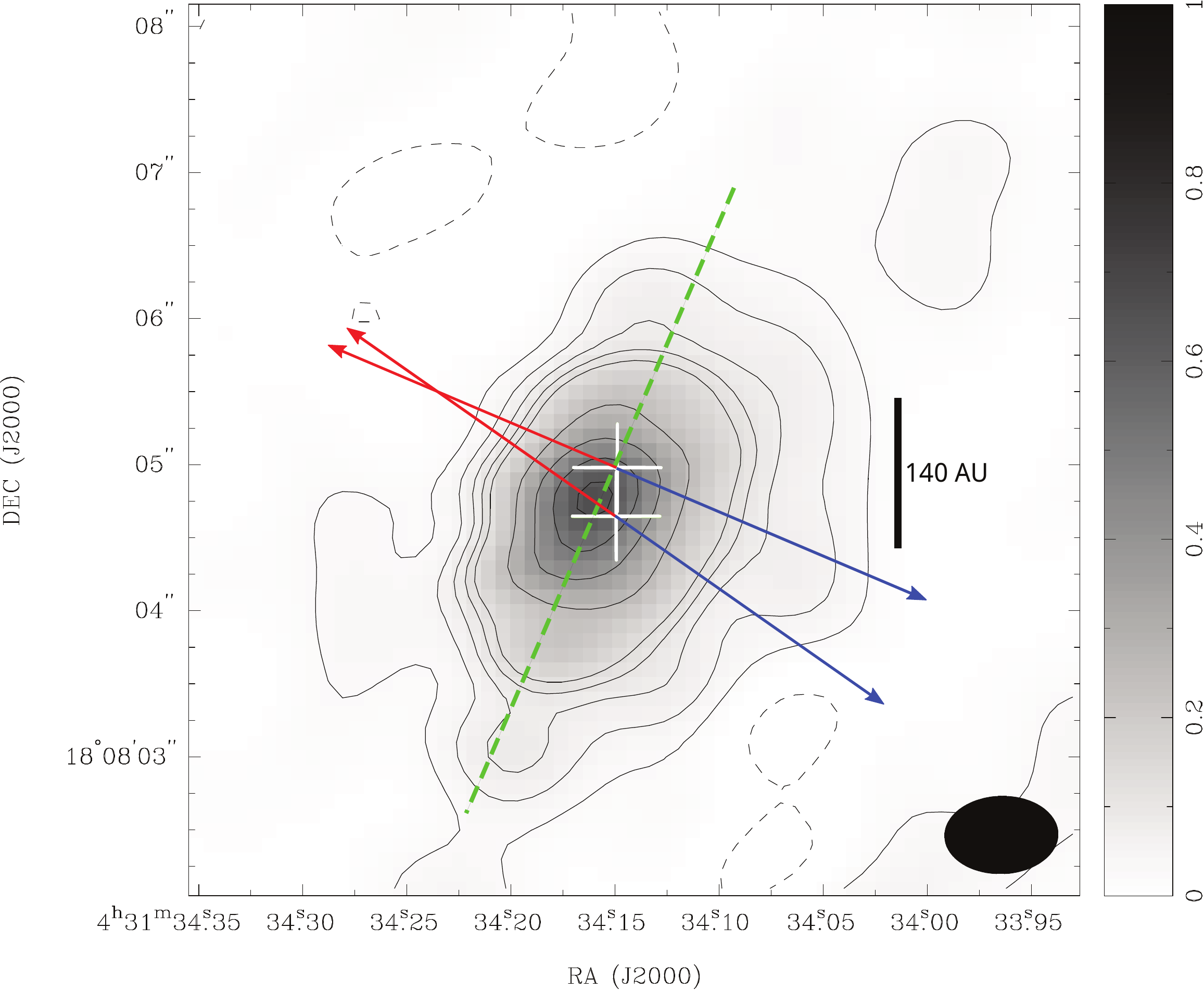}
\caption{343 GHz continuum image of L1551 IRS 5 observed with the SMA. Contour levels are from 3$\sigma$ in steps of 3$\sigma$ until 15$\sigma$, and then in steps of 15$\sigma$ (1$\sigma$ = 8.4 mJy beam$^{-1}$). The highest contour level is 75$\sigma$.
A filled ellipse at the bottom-right corner denotes the synthesized beam size (0$\farcs$77 $\times$ 0$\farcs$53; P.A. = -88$\degr$).
Crosses show the positions of the protobinary in L1551 IRS 5 taken from Lim \& Takakuwa (2006). A green dashed line shows the major axis of the 343 GHz continnum emission as measured from the 2-dimensional Gaussian fitting, and blue and red arrows show the direction of the blueshifted and redshifted jets drived by the protobinary.
\label{smaimages}}
\end{figure}

Figure \ref{smaimages} shows the 343 GHz continuum image of L1551 IRS 5 taken with the SMA.
Crosses in the image show the positions of the protobinary, which is taken from Lim \& Takakuwa (2006).
The continuum emission exhibits an elongated
feature along the northwest to southeast direction, which is approximately
perpendicular to the axes of the associated radio jets (blue and red arrows in
Figure \ref{smaimages}; Rodr\'iguez et al. 2003).
From the two-dimensional Gaussian fitting, the total flux density,
beam-deconvolved size, and the position angle (P.A.) of the continuum emission are derived to be
1.9$\pm$0.3 Jy, 160$\pm$10 AU $\times$ 80$\pm$10 AU, and -23$\degr\pm$10$\degr$,
respectively.
By comparison, the peak 850 $\micron$ flux density observed with JCMT
is 3.16 Jy beam$^{-1}$ at $\sim$14$\arcsec$ resolution \cite{mor06}, and thus the continuum emission detected with the SMA contributes $\sim$60$\%$ of the total continuum flux.
The peak position of the continuum emission derived from the Gaussian fitting
is $\sim$0$\farcs$3 southeast from the position of the northern protostellar component
as measured by Lim \& Takakuwa (2006).
In the L1551 region a global proper motion of $\mu_\alpha=0\farcs012\pm0\farcs002$ yr$^{-1}$
and $\mu_\delta=-0\farcs023\pm0\farcs002$ yr$^{-1}$ has been identified \cite{jon79,rod03},
which implies
the present positions of the protobinary $\sim$+0$\farcs$07$\pm$0$\farcs$01 east
and $\sim$-0$\farcs$14$\pm$0$\farcs$01 south
from the nominal positions measured by Lim and Takakuwa (2006).
The updated position of the northern protostar is still
located $\sim$+0$\farcs$16 north of the continuum peak position.
On the other hand,
the uncertainty of the peak position of the continuum emission measured from the Gaussian fitting is
estimated to be $\sim$0$\farcs$03, much smaller than the offset between the continuum peak
and the protostellar positions.
From these considerations, we suggest that the continuum peak is not associated
with the northern or the southern protostar directly,
and that the continuum emission most likely traces
the flattened circumbinary+circumstellar disks ($e.g.,$ Looney et al. 1997),
embedded in the extended ($\sim$1300 AU) envelope component found with the JCMT observations.
The ratio of the major and minor axes yields the disk inclination angle
from the plane of the sky to be $i=\cos ^{-1}(\frac{\rm 80\pm10AU}{\rm 160\pm10AU})=60\degr\pm5\degr$.

%
%
The mass of the feature detected in the 343 GHz continuum ($\equiv$$M_{d}$) was estimated as
\begin{equation}
M_{d}=\frac{S_{\nu}d^2}{\kappa_{\nu} B_{\nu}(T_d)},
\end{equation}
where $\nu$ is the frequency, $S_{\nu}$ the flux density,
$d$ the distance, $B_{\nu}(T_d)$ the Planck function,
$T_{d}$ the dust temperature, and $\kappa_{\nu}$ the dust opacity per unit gas + dust mass
on the assumption of the gas-to-dust mass ratio of 100.
The frequency dependence of $\kappa_{\nu}$
is expressed as $\kappa_{\nu}$ = $\kappa_{\nu_{0}}$($\nu$/$\nu_{0}$)$^{\beta}$,
where $\beta$ denotes the dust-opacity index. The dust opacity of
Ossenkopf \& Henning (1994) for grains with thin ice mantles coagulated
at a density of $n_{H_{2}}$ $\sim$10$^{6}$ cm$^{-3}$;
$i.e.$, $\kappa_{850~\micron}$ = 0.0182 cm$^{2}$ g$^{-1}$, was adopted.
In L1551 IRS 5 $\beta$ = 1.0 has been measured from the
millimeter and submillimeter continuum data \cite{wu09},
and the measured value of $T_{d}$ ranges from 47 K \cite{mor94} to 94 K \cite{kri12},
resulting in $M_{d}$ = 0.032 -- 0.070 $M_{\odot}$.

\subsection{CS (7--6) Line}

\begin{figure}
\plotone{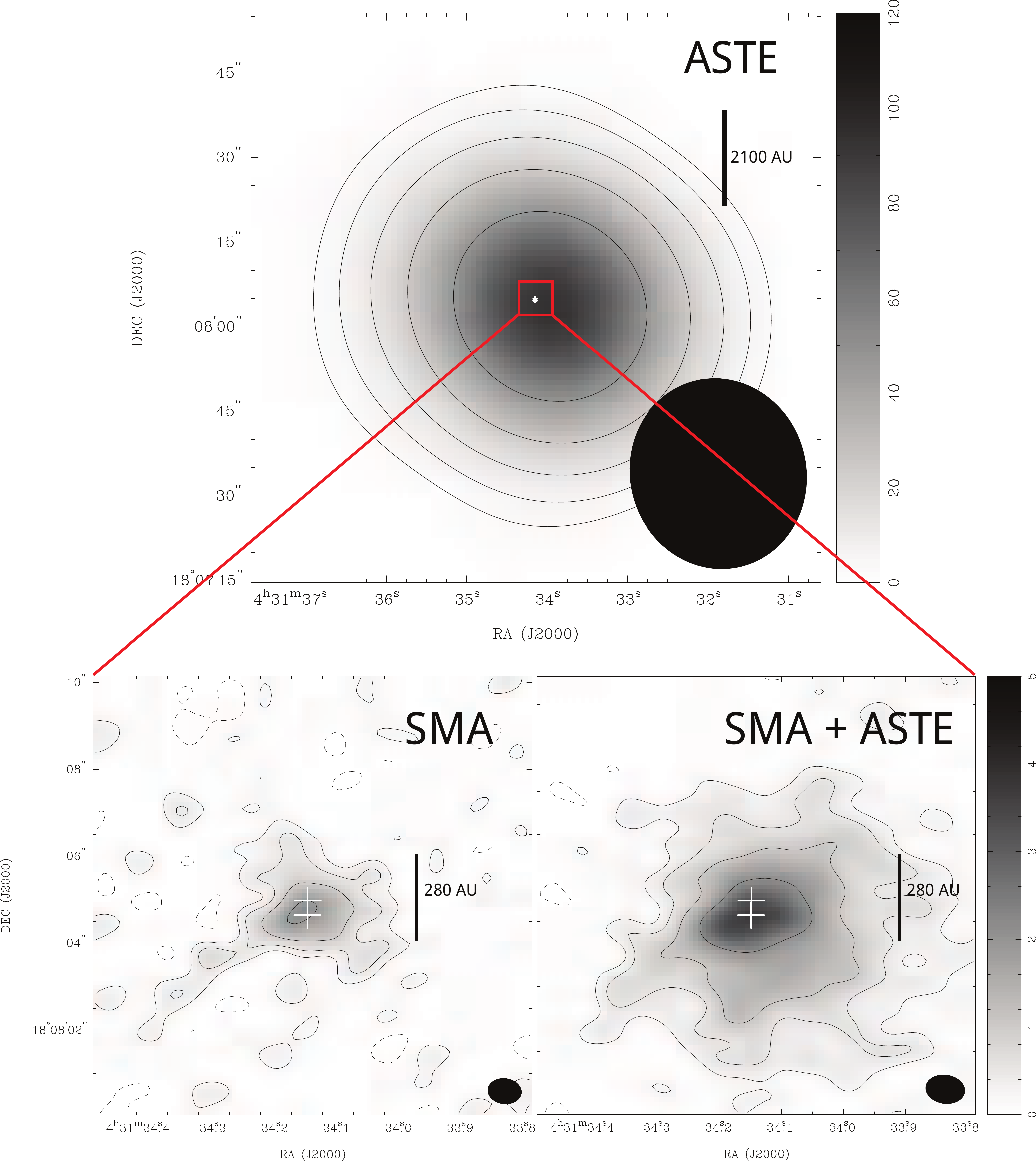}
\caption{Comparison of the ASTE, SMA, and the combined SMA + ASTE moment 0 maps
of the CS (7--6) emission in L1551 IRS 5. The integrated velocity range is
$V_{\rm LSR}$ = 3.8 -- 9.3 km s$^{-1}$.
Contour levels are 2$\sigma$, 4$\sigma$, 8$\sigma$, 16$\sigma$, and 32$\sigma$
(1$\sigma$ = 1.52, 0.122, and 0.133 Jy beam$^{-1}$ km s$^{-1}$ in the ASTE, SMA, and the SMA+ASTE maps).
Filled ellipses at the bottom-right corners of the images
denote the synthesized beam sizes (22$\arcsec$ for the ASTE image, 0$\farcs$77 $\times$ 0$\farcs$58, P.A. = 83$\degr$ for the SMA image, and
0$\farcs$90 $\times$ 0$\farcs$65, P.A. = 82$\degr$ for the SMA + ASTE image.) Crosses show the positions of the protobinary.
\label{mom0}}
\end{figure}

Figure \ref{mom0} compares the ASTE, SMA, and the combined SMA + ASTE moment 0 maps 
of the CS (7--6) emission in L1551 IRS 5. The ASTE image shows an intense blob
centered on the position of the protobinary with a possible extension
toward the southwest ($\sim$2500 AU),
as already discussed by Takakuwa \& Kamazaki (2011).
The SMA-only CS image exhibits a $\sim$300-AU scale, compact feature
centered approximately on the protobinary, which is consistent with our earlier SMA results \citep{tak04}.
The combined SMA + ASTE image recovers the outskirts of the compact feature
seen in the SMA-only image. In our previous study \cite{tak11},
the SMA CS (7--6) data taken as the SMA early science study at a lower angular resolution
($\sim$3$\farcs$2 $\times$ 2$\farcs$0) \cite{tak04} were combined with the ASTE image,
and the combined image shows the central feature associated with the protobinary
plus a weak diffuse component to
the $\sim$1500 AU west and southwest of the protobinary.
In our new combined SMA + ASTE image at a subarcsecond resolution,
the structure of the central component associated with the protobinary is clearly resolved,
while the brightness sensitivity becomes worse than the brightness temperature of the diffuse component.
Hereafter in this paper, we will discuss the velocity structures of the material
associated with the protobinary L1551 IRS 5, as detected
with our new combined SMA + ASTE image cubes.

\begin{figure}
\plotone{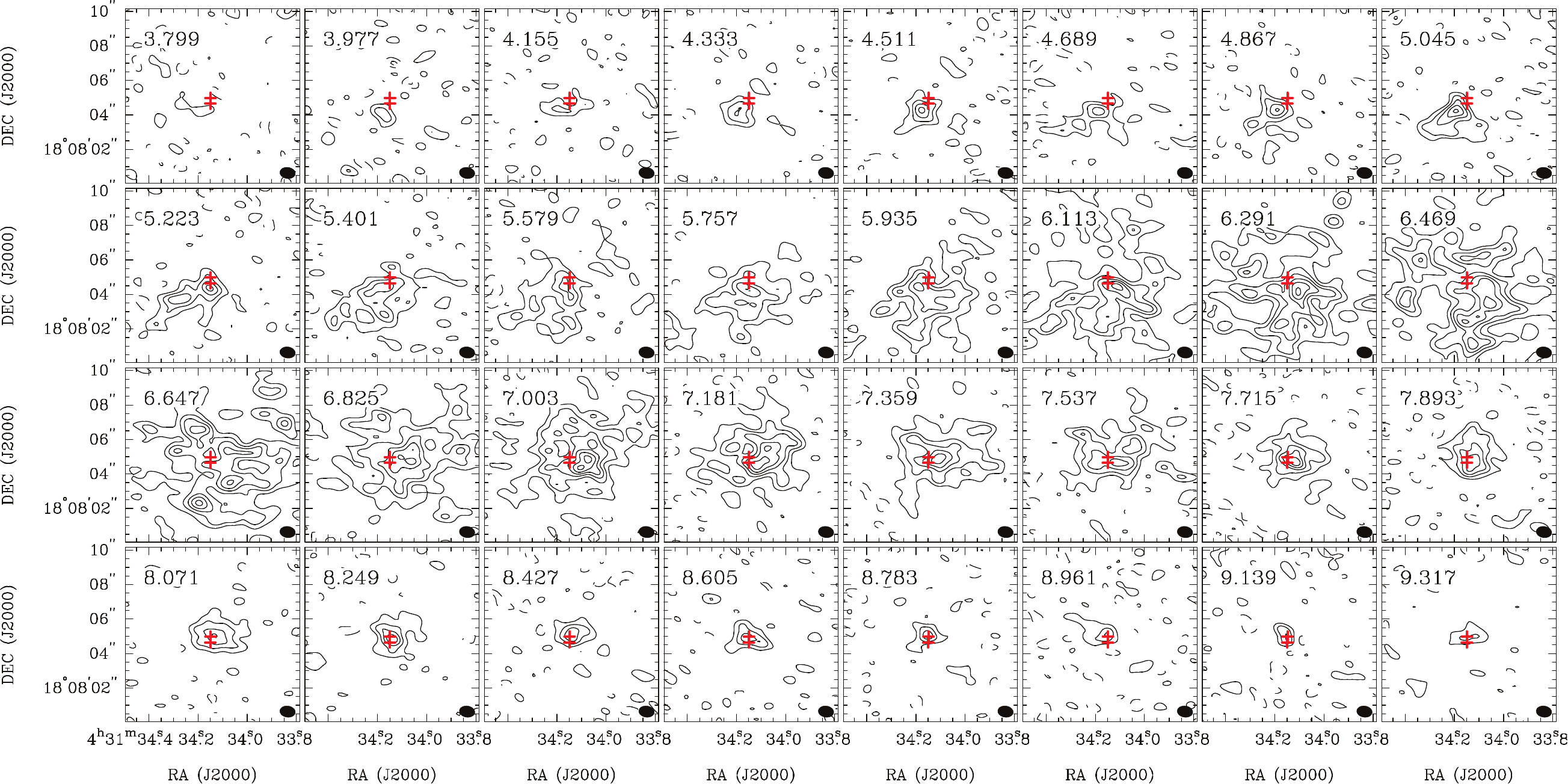}
\caption{SMA + ASTE velocity channel maps of the CS (7--6) line in L1551 IRS 5.
Contour levels are from 2$\sigma$ in steps of 2$\sigma$ (1$\sigma$ = 0.130 Jy beam$^{-1}$).
Crosses indicate the positions of the protobinary,
and a filled ellipse at the bottom-right corner in each panel the synthesized beam
(0\farcs90 $\times$ 0\farcs65; P.A. = 82$\degr$). A number at the top-left corner in each panel denotes the LSR velocity.
\label{chcs}}
\end{figure}

Figure \ref{chcs} shows the SMA + ASTE velocity channel maps of the
CS (7--6) line in L1551 IRS 5. The midpoint between the most extreme velocities
is $v_{\rm LSR}$ = 6.6 km s$^{-1}$, which
is also the symmetric center of the Position-Velocity (P-V) diagrams
(see Section 4).
We therefore adopt $v_{\rm LSR}$ = 6.6 km s$^{-1}$
as the systemic velocity.
In the high-velocity blueshifted region (3.8 km s$^{-1}$ -- 5.0 km s$^{-1}$),
a compact ($\lesssim$2$\arcsec$) CS emission is seen with its peak
to the southeast of the protobinary.
In the low-velocity blueshifted region (5.2 km s$^{-1}$ -- 6.1 km s$^{-1}$),
an extended CS emission with multiple peaks is seen, and the emission centroids
appear to be located to the south of the protobinary.
In the velocities close to the systemic velocity (6.3 km s$^{-1}$ -- 6.8 km s$^{-1}$),
both the blueshifted and redshifted emission are extended in the entire region.
In the low-velocity redshifted region (7.0 km s$^{-1}$ -- 7.9 km s$^{-1}$),
an intense emission peak is seen to the west of the protobinary, whereas
the overall emission centroids appear to be located to the north of the protobinary,
which is opposite to the low-velocity blueshifted region.
In the highly redshifted region (8.1 km s$^{-1}$ -- 9.3 km s$^{-1}$),
a compact CS emission, just as in the case of the high-velocity blueshifted region,
is seen to the north of the protobinary.

\begin{figure}
\plotone{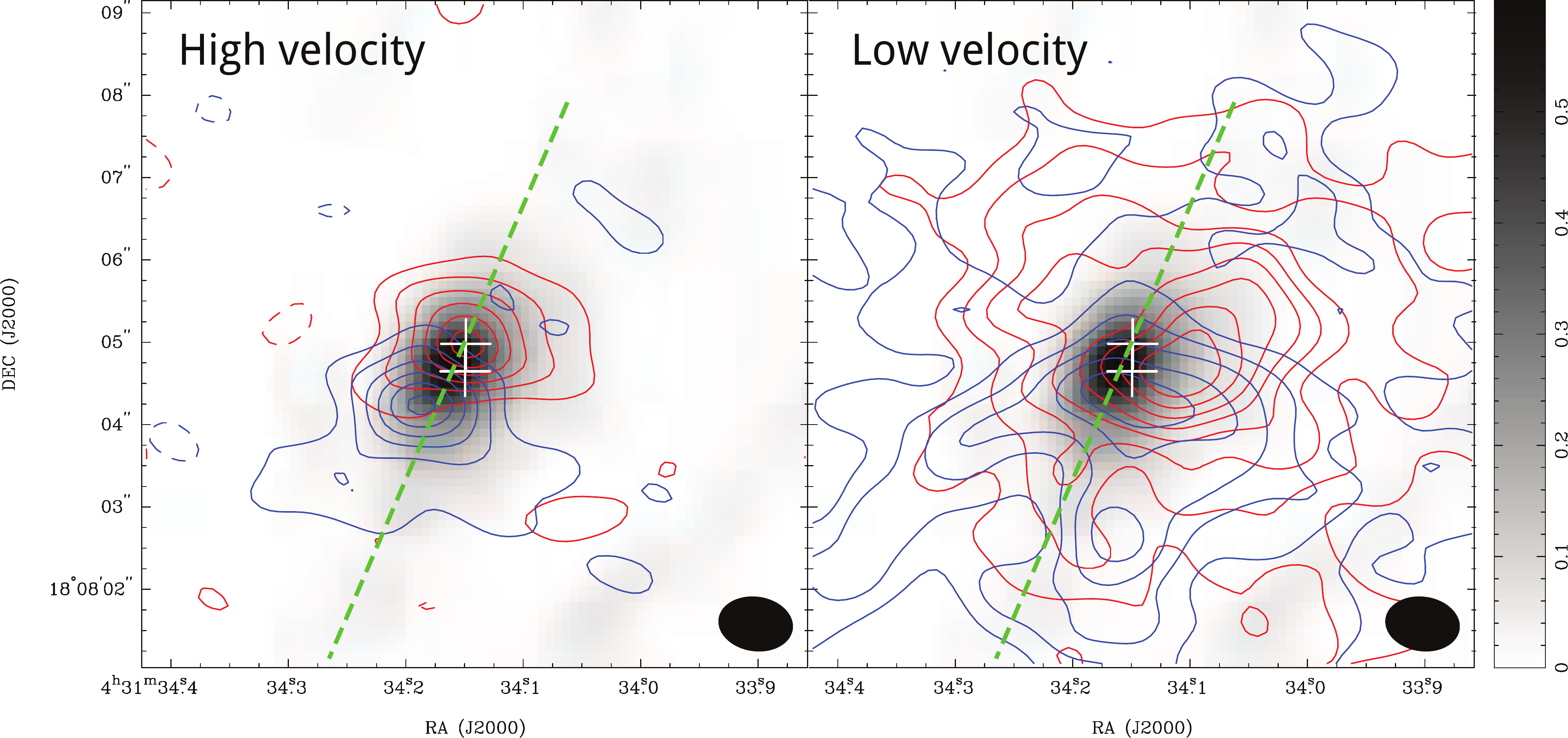}
\caption{Distributions of the high- and low-velocity blueshifted (\emph{blue contours}) and redshifted (\emph{red}) CS (7--6) emission
in L1551 IRS 5 obtained with the SMA and ASTE, superposed on the 343 GHz continuum image (\emph{grey scale}). Contour levels are from 3$\sigma$ in steps of 3$\sigma$. The 1$\sigma$ rms noise levels of the high-velocity blueshifted, redshifted, low-velocity blueshifted, and redshifted emission are 0.065, 0.077, 0.065, 0.057 Jy beam$^{-1}$ km s$^{-1}$, respectively, and the integrated velocity ranges are 3.799 km s$^{-1}$ -- 5.045 km s$^{-1}$, 8.071 km s$^{-1}$ -- 9.495 km s$^{-1}$, 5.223 km s$^{-1}$ -- 6.469 km s$^{-1}$, and 6.647 km s$^{-1}$ -- 7.893 km s$^{-1}$, respectively.
Crosses show the positions of the protobinary. A filled ellipse at the bottom-right corner in each panel denotes the synthesized beam
(0\farcs90 $\times$ 0\farcs65; P.A. = 82$\degr$). Dashed lines show the major axis of the 343 GHz continuum emission.
\label{brcs}}
\end{figure}

To better highlight the spatial-kinematic distributions of the CS emission, in Figure \ref{brcs} we show
the integrated-intensity maps in the high-velocity and low-velocity blueshifted and redshifted ranges separately,
superposed on the 343 GHz dust continuum image.
In the high-velocity range, the blueshifted and redshifted CS emission peaks are symmetrically
located to the southeast and northwest with respect to the continuum peak.
The position angle of the axis connecting the blueshifted and redshifted emission peaks
matches that of the major axis of the 343 GHz continuum emission (dashed green line
in Figure \ref{brcs}) within $\sim$10$\degr$,
and the separation between the two peaks ($\sim$140 AU) is similar to the FWHM size
of the 343 GHz continuum emission ($\sim$160 AU).
Thus, the high-velocity CS emission exhibits a southeast (blue) to northwest (red)
velocity gradient along the major axis of the circumbinary material as seen in the 343 GHz
dust-continuum emission.
In contrast, in the low-velocity range, the blueshifted and redshifted CS emission 
are extended and overlap with each other.
The blueshifted emission peak is located to the south of the protobinary,
while the redshifted emission peak is located to the west of the protobinary.

\begin{figure}
\plotone{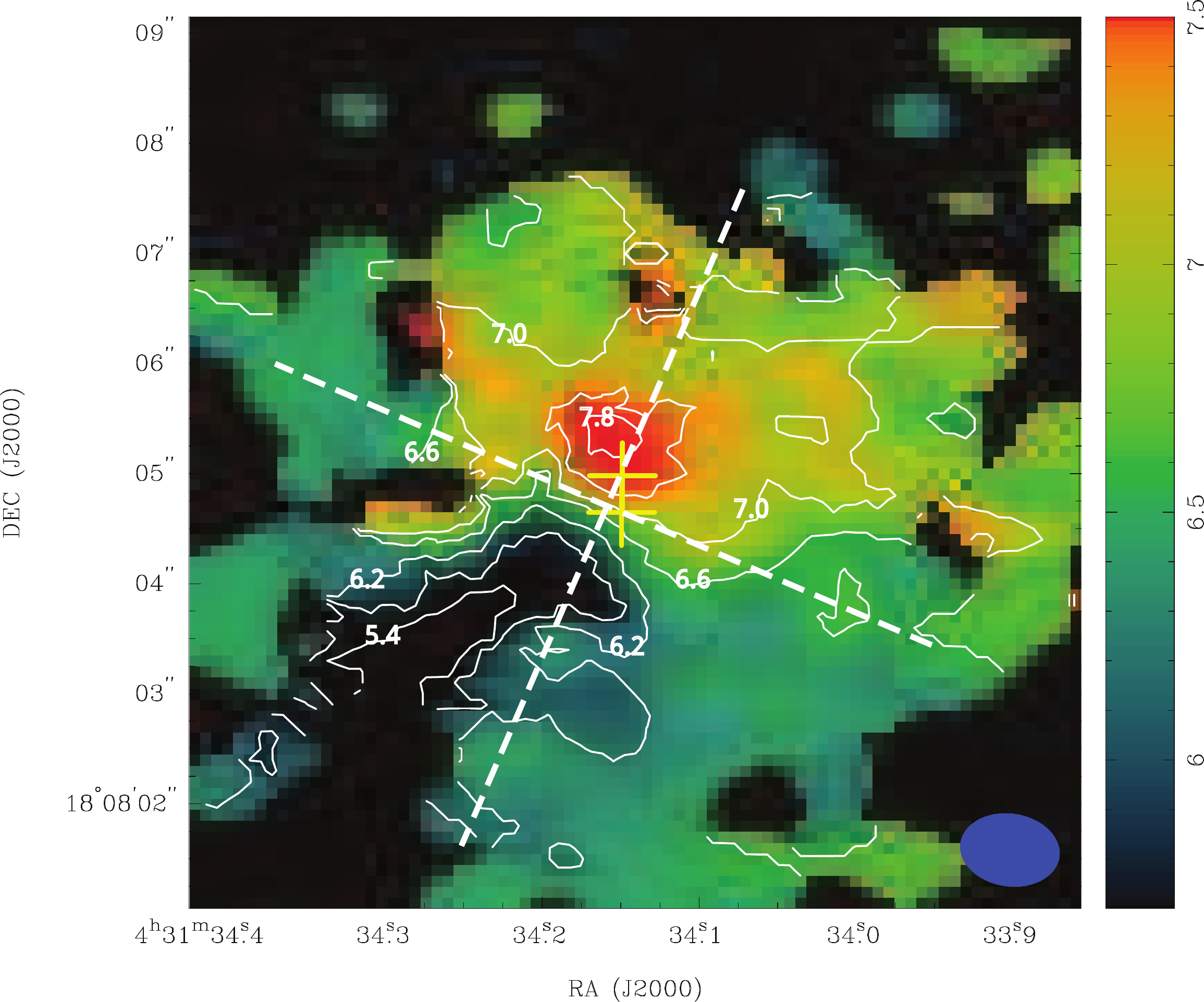}
\caption{SMA + ASTE moment 1 map of the CS (7--6) emission in L1551 IRS 5.
Yellow crosses show the positions of the protobinary, and white dashed lines the major and minor axes of the 343 GHz continuum emission.
Contour levels are in steps of 0.4 km s$^{-1}$, and the bluest and reddest contour levels to the south-east and north-west of the protobinary
are 5.4 km s$^{-1}$ and 7.8 km s$^{-1}$, respectively. Some of the contours are also labeled with their values. A filled ellipse at the bottom-right corner denotes the synthesized beam size (0\farcs90 $\times$ 0\farcs65; P.A. = 82$\degr$).
\label{mom1}}
\end{figure}

Figure \ref{mom1} shows the intensity-weighted mean-velocity map of the CS emission
in L1551 IRS 5. Along the major axis of the 343 GHz continuum emission
(tilted vertical dashed line in Figure \ref{mom1}), the CS emission to the north of
the protobinary is redshifted with the most redshifted velocity of 7.8 km s$^{-1}$
at $\sim$0.7$\arcsec$ offset from the continuum peak,
while the emission to the south is blueshifted with the most blueshifted velocity of 5.4 km s$^{-1}$
at $\sim$1.5$\arcsec$.
This velocity feature reflects the velocity gradient seen in the high-velocity CS emission
as shown in Figure \ref{brcs}. The global feature of the northern redshifted
and southern blueshifted emission extends to the area outside of the continuum emission.
We will discuss these velocity features with our simple models
in the next section.

\section{Analyses}

As shown in the previous section, the CS (7--6) emission surrounding L1551 IRS 5
exhibits distinct high-velocity and low-velocity features.
The high-velocity emission comprises blueshifted and redshifted components
located to the southeast and northwest, respectively,
and as a whole the emission extent ($\sim$160 AU) and the position angle
($\sim$$-30\degr$) match with those of the compact continuum emission.
The low-velocity components extend beyond the continuum-emission region
up to $\sim$1000 AU, and the blueshifted and redshifted emission overlap with each other.
Here, with our simple kinematical models
we will interpret these results in the high-velocity and low-velocity regions separately.
We adopt the peak position of the continuum emission as derived from the 2-dimensional
Gaussian fitting as the center of the mass in these models.

\subsection{Central Compact CS Component: Keplerian Disk?}

\begin{sloppypar}
The high-velocity CS (7--6) emission immediately surrounding L1551 IRS 5 exhibits a
clear velocity gradient along the northwest (red) -- southeast (blue) direction.
Since the direction of the velocity gradient is parallel
to the major axis of the dust-continuum emission and 
the high-velocity CS emission traces the overall structure of the dust-continuum emission,
a natural interpretation of the high-velocity CS emission is a rotating circumstellar disk.
Recent interferometric observations of protostellar sources have been successfully modeling
the circumstellar disks as geometrically-thin Keplerian disks \cite{tob12,tak12,tak13,yen13,yen14,sak14}.
Recent theoretical simulations of disk formation have also shown that
at $\sim$100--200 AU scales the vertical extents of the disks are negligible
($h\sim$20 AU; Machida et al. 2011b).
L1551 IRS 5 is associated with a 2500-AU scale flattened envelope
seen in the C$^{18}$O (1--0) emission, with the position angle ($\theta \sim$$-20\degr$)
and inferred inclination angle ($i \sim$$60\degr$) similar to those of
the compact ($\sim$160 AU) dust-continuum emission found
with the present SMA observations.
Thus, we performed model fittings of a geometrically-thin Keplerian disk to the
high-velocity CS emission.
\end{sloppypar}

The procedure of the model fittings is basically the same as that adopted by Takakuwa et al. (2012).
Since we assume that the disk is geometrically thin, the
line-of-sight velocity at each position in right ascension
and declination with respect to the disk center ($\equiv v_{LOS} \left( \alpha, \delta \right)$)
can be expressed as
\begin{equation}
v_{LOS} ( \alpha, \delta) = v_{sys} + v_{rot} \left( r \right) \sin i \cos \left( \phi - \theta \right) + v_{rad} \left( r \right) \sin i \sin \left( \phi - \theta \right),
\end{equation}
where
\begin{equation}
r = \sqrt{ \left( \frac{x}{ \cos i} \right) ^2 + y^2},
\end{equation}
\begin{equation}
x = \alpha \cos \left( \theta \right) - \delta \sin \left( \theta \right),
\end{equation}
\begin{equation}
y = \alpha \sin \left( \theta \right) + \delta \cos \left( \theta \right).
\end{equation}
In the above expressions, $v_{sys}$ is the systemic velocity, $\theta$ is the
position angle of the disk major axis, 
$\phi$ is the azimuthal angle from the major axis on the disk plane, $i$ is the
disk inclination angle from the plane of the sky, $\alpha$ and $\delta$ are
coordinates along right ascension and declination with respect to the disk center,
and $x$ and $y$ are coordinates along the minor and major axes of the disk, respectively. $v_{rot}(r)$
denotes the rotational velocity of the disk as a function of
the radius ($\equiv r$), and $v_{rad}(r)$ is the radial velocity.
In the case of Keplerian rotation, $v_{rad}(r)=0$ and
$v_{rot}(r)$ is expressed as
\begin{equation}
v_{rot} \left( r \right) = \sqrt{ \frac{ GM_\star}{r}},
\end{equation}
where $G$ is the gravitational constant and $M_\star$ is the mass of the central star.
The 3-dimensional flux map of the CS emission, $i.e.$,
the velocity channel maps, of
the model disk ($\equiv S_{model} \left( \alpha, \delta, v \right)$)
can be expressed as
\begin{equation}
S_{model} \left( \alpha, \delta, v \right) = \left( S_{mom0} \left( \alpha, \delta \right) / \sigma \sqrt{ 2 \pi} \right) \times \exp \left( \frac{ - \left( v - v_{LOS} \left( \alpha, \delta \right) \right)^2}{ 2.0 \sigma^2} \right),
\end{equation}
where $S_{mom0} \left( \alpha, \delta \right)$ denotes the moment 0 map of the
model disk and $\sigma$ the internal velocity dispersion.

In our fitting, the moment 0 map of the model disk is assumed to be the same as
the real observed moment 0 map (Figure \ref{mom0} bottom-right).
The purpose of our modeling is to investigate the velocity structure of the high-velocity
CS emission, and to reproduce the observed velocity channel maps
with kinematical and geometrical parameters.
Even if the model moment 0 map is assumed to be the same as the observed moment 0 map,
different kinematical parameters yield different velocities where the CS emission arises,
and thus different model velocity channel maps.
On the other hand, modeling the moment 0 map itself requires parameters of the profiles of surface density, temperature, and the molecular abundance, that is, parameters not related to the velocity structure. Therefore, to discuss the nature of the velocity structure our approach should be sufficient.
To reduce the number of the fitting parameters, 
we fix the disk inclination angle $i = -60 \degr$
derived from the ratio of the extents along the major and minor axes of the continuum emission,
the systemic velocity of $v_{sys}$ = 6.6 km s$^{-1}$, and the
velocity dispersion $\sigma$ = 0.7 km s$^{-1}$ inferred from the inspection of the
observed CS velocity channel maps.
Only the high-velocity parts of the CS velocity channel maps
(3.8 -- 5.0 km s$^{-1}$ and 8.1 -- 9.3 km s$^{-1}$), which trace
the structure of the continuum emission (see Figure \ref{brcs} $left$), were included
in the fitting.
On these assumptions, we conducted minimum $\chi^2$-fittings of the
velocity channel maps of the model Keplerian disk to
the observed CS velocity channel maps, with $M$ and $\theta$ as fitting parameters;
\begin{equation}
\chi^2 = \sum_{ \alpha, \delta, v} \left( \frac{ S_{obs} \left( \alpha, \delta, v \right) - S_{model}^{M_\star, \theta} \left( \alpha, \delta, v \right)}{ \sigma_{rms}} \right)^2 \Bigg / \sum_{ \alpha, \delta, v},
\end{equation}
where $S_{obs} \left( \alpha, \delta, v \right)$ denotes the observed velocity channel maps,
$S_{model}^{M_\star, \theta}$ the model velocity channel maps with given
$M_\star$ and $\theta$, and $\sigma_{rms}$ the rms noise level of the
observed velocity channel maps.

\begin{figure}
\plotone{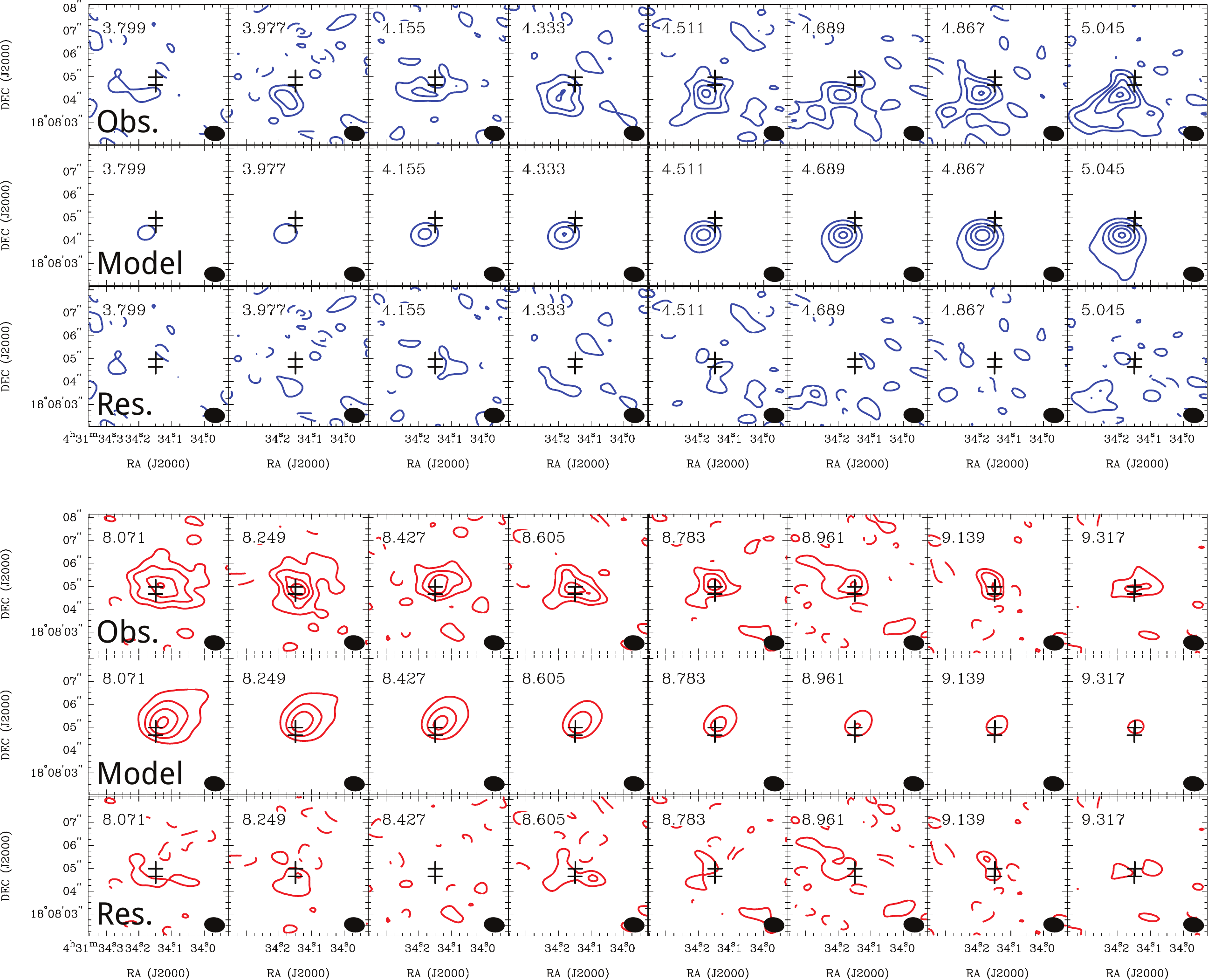}
\caption{Best-fit results of the $\chi^2$ model fitting of the geometrically-thin Keplerian disk to the SMA+ASTE CS (7-6) velocity channel maps in L1551 IRS 5 at the highly blueshifted (\emph{blue contours}) and redshifted (\emph{red})  velocities. Upper, middle, and lower panels show the observed, model, and the residual velocity channel maps, respectively, where the best-fit parameters are $M_\star=0.5$ $M_\odot$ and disk position angle $\theta=-33^\circ$. Contour levels are from 2$\sigma$ in steps of 2$\sigma$ ($1\sigma=0.130$ Jy beam$^{-1}$). Crosses show the positions of the protobinary, and the filled ellipses at the bottom-right corners the SMA + ASTE synthesized beam ($0\farcs90\times 0\farcs65; P.A.=-82^\circ$).
\label{modelfit}}
\end{figure}

After running $\sim$200 sets of different parameters,
we obtained the best-fit Keplerian-disk model,
in which $M_\star$ = $0.5^{+0.6}_{-0.2}$ $M_\odot$
(which is equivalent to $v_{rot}$ = $1.7^{+0.9}_{-0.3}$ km s$^{-1}$
at a radius $r$ = 140 AU) and $\theta = -33^{+17}_{-24}$ degrees.
The error bars correspond to the $\chi^2+1$ values.
The middle and lower panels of Figure \ref{modelfit} show the velocity channel maps
of the best-fit model and the residuals, respectively.
Since the moment 0 map of the model Keplerian disk is assumed
to be the same as the real observed moment 0 map, the model velocity channel maps
do not exhibit
perfectly smooth and symmetric features as expected from normal disk models.
The residual velocity channel maps do not show any systematic
features, and the rms of the residual maps is $\sim$0.147 Jy beam$^{-1}$,
which is close to the noise level of the observed velocity channel maps ($\sim$0.130 Jy beam$^{-1}$).
The derived position angle of the Keplerian disk ($\theta = -33\degr$)
is consistent with the major axis of the 343 GHz continuum emission ($\theta = -23\degr$),
and approximately perpendicular to the axes of the jets driven by the protobinary.
We also fixed $\theta = -23\degr$ and performed another model fitting with
$M_\star$ as the only fitting parameter, and found that the best fit value of $M_\star$ is still the same,
with a slightly larger rms of the residual channel maps ($\sim$0.149 Jy beam$^{-1}$).
As will be discussed in the next subsection,
the radius of the 343 GHz continuum emission ($\sim$160 AU) is almost the same
as the expected centrifugal radius derived from the outer envelope rotation
and the central stellar mass derived from the present fitting.
Furthermore,
in the high-velocity regions of the fitting, the SMA-only image cube recovered at least $\gtrsim$64\%
of the ASTE flux.
The model fitting to the SMA-only image cube
in the same high-velocity regions provides a similar fitting result, with inclination angle $i$ also as a fitting parameter
($M_\star\sim$0.7 $M_{\odot}$, $\theta\sim$$-30\degr$, and $i\sim$$-44\degr$).
These results show that the observed 343 GHz continuum emission and high-velocity CS emission
are consistent with a Keplerian disk surrounding the L1551 IRS 5 binary system,
or Keplerian circumbinary disk as in the case of L1551 NE \cite{tak12,tak13}.
Interestingly, the estimated position and inclination angles of the circumbinary
disk around L1551 IRS 5 are similar to those of the circumbinary
disk around L1551 NE,
but their rotations are in the opposite directions (in L1551 NE the northern part is blueshifted).

\subsection{Extended Component: Infalling Envelope\\
with a Reduced Infalling Velocity?}

While the high-velocity blueshifted and redshifted CS emission (Figure \ref{brcs} $left$)
are well reproduced by the Keplerian disk model, the low-velocity blueshifted and redshifted
CS emission (Figure \ref{brcs} $right$) appear to be distinct from the inner Keplerian disk.
Momose et al. (1998) have found a $\sim$2500-AU scale flattened envelope
around L1551 IRS 5 in the C$^{18}$O (1--0) emission with the NMA,
which exhibits infalling and rotating gas motion with a conserved specific angular momentum.
The rotational velocity with a conserved specific angular momentum is expressed as,
\begin{equation}
v_{rot} \left( r \right) = \frac{j}{r},
\end{equation}
where $j$ denotes the specific angular momentum. Momose et al. (1998) measured
$j$ $\sim$168 AU km s$^{-1}$ in the flattened C$^{18}$O envelope around L1551 IRS 5.
Equating eq. (6) and eq. (9) provides the transitional radius from the outer infalling envelope
to the inner Keplerian disk. For $M_{\star}$ = 0.5 $M_\odot$ and
$j$ = 168 AU km s$^{-1}$, the outermost Keplerian radius is calculated to be $r_{kep}$ $\sim$64 AU
or 128 AU in diameter,
which is almost the same as the FWHM size of the continuum emission ($\sim$160 AU).
This means that the outer rotation of the infalling envelope seen in the C$^{18}$O (1--0) emission
naturally connects to the rotation of the Keplerian circumbinary disk as traced by the high-velocity CS emission
and the dust-continuum emission.
The low-velocity CS (7--6) emission we observed with the SMA and ASTE
may trace the innermost part of this infalling envelope.

\begin{figure}
\plotone{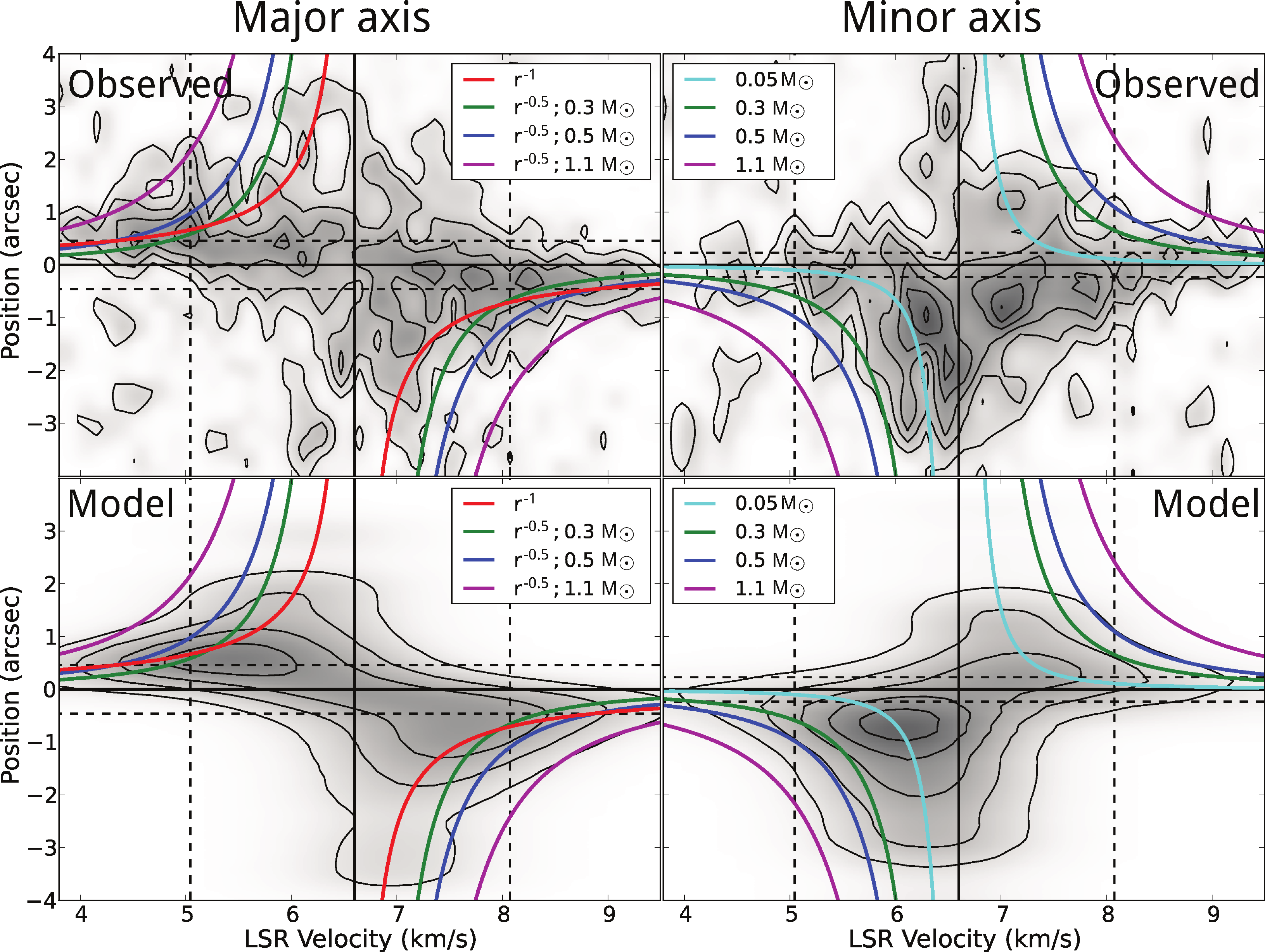}
\caption{SMA+ASTE (\emph{upper panels}) and model (\emph{lower}) Position -- Velocity (P-V) diagrams of the CS (7--6) emission in L1551 IRS 5
along the major (P.A. = 147$\degr$; $Left$) and minor axes (P.A. = 57$\degr$; $Right$) of
the Keplerian circumbinary disk.
Contour levels are from 2$\sigma$ in steps of 2$\sigma$ (1$\sigma$ = 0.130 Jy beam$^{-1}$).
Horizontal thick and dashed lines denote the centroid position of the 343 GHz continuum emission and the size of the the Keplerian disk (64 AU in halfwidth along the major axis and 32 AU along the minor axis), respectively. Vertical thick and dashed lines denote the systemic velocity (= 6.6 km s$^{-1}$) and the velocity borderline used in Figure \ref{brcs} to separate the high-velocity and low-velocity components, respectively.
In the left panels, the red curves show the rotation curve with a conserved angular momentum ($j$ = 168 AU km s$^{-1}$); the green, blue, and purple curves show the Keplerian rotation curves with the central stellar masses of 0.3, 0.5, and 1.1 $M_{\odot}$, respectively.
In the right panels, cyan, green, blue, and purple curves show free-fall gas motions with the central stellar masses of
0.05, 0.3, 0.5, and 1.1 $M_{\odot}$, respectively.
\label{pv}}
\end{figure}

Figure \ref{pv} (upper panels) shows the observed Position-Velocity (P-V) diagrams of
the CS (7--6) emission in L1551 IRS 5 along the major ($left$) and minor axes ($right$)
of the Keplerian disk identified by the high-velocity CS emission.
In the P-V diagrams, horizontal dashed lines delineate the size of the Keplerian disk,
and vertical dashed lines the velocity borderlines used to separate the high-velocity
and low-velocity components.
In the observed P-V diagram along the major axis, the velocity gradient along the southeast (blueshifted)
to the northwest (redshifted) direction is evident both within and outside of the Keplerian-disk region.
The lower-velocity emission outside of the Keplerian disk extends to
$r$ $\sim$500 AU, while the higher-velocity emission originated from the Keplerian disk
is located closer to the central protobinary. To contrast the velocity feature
of the outer lower-velocity emission with that of the inner Keplerian disk,
curves of two kinds of rotations are drawn in the P-V diagram along the major axis.
One is the rotation with the conserved angular momentum
derived from the C$^{18}$O envelope by Momose et al. (1998)
($i.e.$, eq. (9) with $j$ = 168 AU km s$^{-1}$; red curves in Figure \ref{pv}).
The other is the Keplerian rotation derived in the previous subsection
($i.e.,$ eq. (6) with $M_\star$ = $0.5^{+0.6}_{-0.2}$ $M_\odot$;
green, blue, and purple curves in Figure \ref{pv}).
Although the Keplerian and $r^{-1}$ rotation curves are almost indistinguishable
in the higher-velocity parts,
the $r^{-1}$ rotation curve with $j$ = 168 AU km s$^{-1}$, which reproduces
the rotation of the C$^{18}$O envelope, appears to better trace the velocity
feature of the lower-velocity CS emission. Hereafter we fix $j$ = 168 AU km s$^{-1}$
for the rotation traced by the low-velocity CS emission.

In the P-V diagram along the minor axis, the higher-velocity parts do not exhibit
any clear velocity gradient, as expected from the Keplerian rotation.
On the other hand, the lower-velocity CS emission shows that
in the northeastern part the CS emission is primarily redshifted while
in the southwestern part both the blueshifted and redshifted CS emission are seen.
Thus, there is a possible northeast (red) to southwest (blue) velocity gradient
in the lower-velocity CS emission.
Since the associated bipolar outflows are also blueshifted to
the southwest and redshifted to the northeast \cite{mor06,wu09},
a possible interpretation for the low-velocity components is a molecular outflow.
If the low-velocity CS components indeed trace the associated molecular outflow,
then based on their projected velocity with respect to
the systemic velocity and the projected separation from the protobinary 
along the minor axis ($\lesssim$0.5 km s$^{-1}$ and $\sim$1$\arcsec\sim$140 AU, respectively),
and assuming that the outflow emerges perpendicular to the Keplerian disk,
the deprojected outflow velocity is estimated to be
$v_{flow}$ $\sim$0.5 km s$^{-1}$/$\cos i$ $\sim$1 km s$^{-1}$,
and the deprojected separation $l_{flow}$ $\sim$140 AU/$\sin i$ $\sim$162 AU.
By comparison, at a given separation from the central gravity source
the escape velocity ($\equiv v_g$) is
\begin{equation}
v_g = \sqrt{\frac{2GM_{\star}}{l_{flow}}}.
\end{equation}
Since $M_{\star}$ is derived to be 0.5 $M_{\odot}$ from the Keplerian rotation,
we find $v_g\sim$2.3 km s$^{-1}$, a factor of 2 higher than the deprojected outflow
velocity estimated above ($v_{flow}\sim$1 km s$^{-1}$).
The low-velocity components are therefore bound to the protostellar system.
Furthermore, contrary to expectations for a molecular outflow driven along
the northeast to southwest direction, the low-velocity components are widely distributed
and do not show any bipolar feature.
Thus, it is unlikely that the low-velocity components
originate from the associated molecular outflow.

\begin{figure}
\plotone{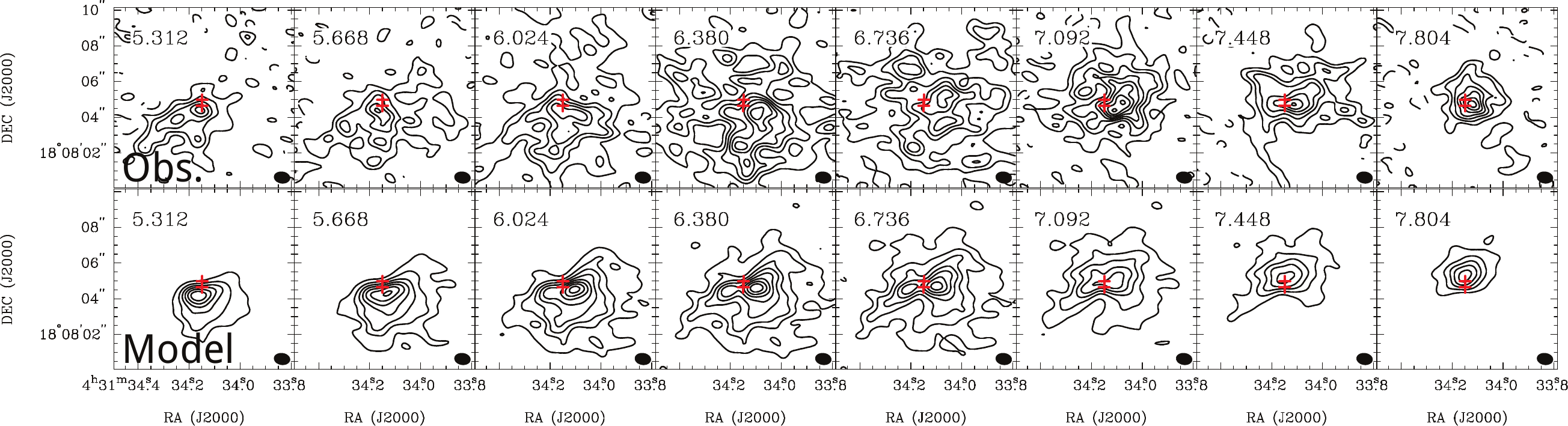}
\caption{Comparison of the observed SMA+ASTE velocity channel maps of the CS line (\emph{upper row}) to the velocity channel maps of the geometrically-thin infalling and rotating envelope model (\emph{lower row}) at the lower velocities ($-1.377$ km s$^{-1}$ $\sim$ $0.937$ km s$^{-1}$ from the systemic velocity). Contour levels are from 2$\sigma$ in steps of 2$\sigma$ (1$\sigma$ = 0.092 Jy beam$^{-1}$). Details of the model are described in Section 4.2. Crosses show the position of the protobinary, and filled ellipses at the bottom-right corners indicate the SMA + ASTE synthesized beam ($0\farcs90\times 0\farcs65$; P.A. = $-82^\circ$).
\label{figX}}
\end{figure}

Instead, the northeast (red) to southwest (blue) velocity gradient
in the low-velocity CS emission may trace the infalling motion
toward the central Keplerian disk in the flattened protostellar envelope.
Since the associated bipolar outflows are blueshifted to
the southwest and redshifted to the northeast \cite{mor06,wu09}, the near-side of the flattened
envelope is northeast and the far-side southwest, and the redshifted emission on the near-side
and the blueshifted emission on the far-side imply infalling motion in the envelope.
Assumption of the flattened morphology of the CS emission is probably valid,
since at $r \sim$500 AU the expected scale height of the flared disk / envelope is
$\lesssim$50 AU \cite{mac14}, and Momose et al. (1998) successfully modeled
the $\sim$2500-AU scale envelope seen in the C$^{18}$O (1--0) emission as
a flattened infalling and rotating envelope.
To quantitatively understand the infalling motion,
in the P-V diagram along the minor axis we draw curves expected from the free-fall motion
expressed as;
\begin{equation}
v_{rad} \left( r \right) = \sqrt{ \frac{ 2GM_\star}{r}}.
\end{equation}
From our Keplerian model fitting to the high-velocity CS emission, the central stellar
mass is estimated to be $\sim0.5^{+0.6}_{-0.2}$ $M_{\odot}$. The blue, purple, and green curves in Figure \ref{pv}
show the infalling curves with the central stellar mass derived from the
Keplerian fitting and its upper and lower ends,
respectively.
It is evident that the free-falling curves with the best-fit value $M_\star$= 0.5 $M_\odot$,
and even with the lower end ($M_\star$= 0.3 $M_\odot$), do not 
trace the emission ridge in the P-V diagram along the minor axis.
If, for example, we reduce the central stellar mass by one order of magnitude to 0.05 $M_\odot$,
the infalling curves (cyan curves in Figure \ref{pv}) trace the bulk of the CS emission ridge
much better.
These results show that the infalling motion, if any,

To reproduce the low-velocity CS emission as the rotating and infalling flattened envelope
with the same rotational profile as that of the outer C$^{18}$O envelope and
the slower inward motion than the free-fall motion, we have generated a toy model
of a geometrically-thin disk with velocity fields described by
\begin{eqnarray}
& v_{rot} \left( r \right) = \displaystyle\sqrt{\frac{GM_\star}{r}}, v_{rad} \left( r \right) = 0 & \left( \textrm{Keplerian region} \right), \\
& v_{rot} \left( r \right) = \displaystyle\frac{j}{r}, v_{rad} \left( r \right) = \displaystyle\sqrt{\frac{2GM_\star^{inf}}{r}} & \left( \textrm{infalling region} \right),
\end{eqnarray}
where $M_\star^{inf}$ denotes the reduced central stellar mass to reproduce the infalling velocity.
As in the case of the Keplerian-disk fitting described in the last subsection,
the two-dimensional intensity distribution of the model disk
was taken to be the same as the observed CS moment 0 map.
Asymmetric or clumpy features in the model velocity channel maps or P-V diagrams
thus originate from the observed CS moment 0 map.
To separate the inner Keplerian region from the outer infalling region,
we defined the area within the FWHM of the Gaussian fit of the continuum emission
as the Keplerian region, and the area outside as the infalling region.
Based on the results of the $\chi^2$ model fitting to the high-velocity CS emission,
we adopted $M_\star=0.5$ $M_\odot$, $v_{sys}=6.6$ km s$^{-1}$, $i=60\degr$, $\theta=-33\degr$,
and $\sigma_{gas}=0.7$ km s$^{-1}$.
The specific angular momentum
$j=168$ AU km s$^{-1}$ and $M_\star^{inf}$ = 0.05 $M_\odot$ are also adopted from previous discussions.
With these parameters, we created the model velocity channel maps with the same procedure
as that described in the last subsection, and then the model P-V diagrams, which are
shown in Figure \ref{pv} ($lower~panels$).
The model P-V diagrams well reproduce the observed velocity structures.
Along the major axis, the model P-V diagram shows both the high-velocity Keplerian
rotation part and the low-velocity envelope part with the $r^{-1}$ rotation.
The model P-V diagram along the minor axis shows that there is no velocity gradient in the high-velocity Keplerian part,
while in the low-velocity part the velocity gradient caused by the slow infalling motion is evident.
Figure \ref{figX} compares the observed and model velocity channel maps in the low-velocity region.
For the velocities of 5.3 -- 6.0 km s$^{-1}$ and 7.1 -- 7.8 km s$^{-1}$, the positions of the emission centroids
of the model channel maps
are located to the south and north of the protobinary, respectively, and the position angle of the axis
connecting between the blueshifted and redshifted components is $\sim$0$\degr$, different from
the position angle of the major axis of the high-velocity Keplerian disk component. This is because
both the rotational and infalling motions in the envelope contribute to the observed velocities, as expressed in eq. (2).
Although the observed velocity channel maps show multiple emission peaks, the overall trend of
the emission centroids in these velocities is consistent with that of our model.
Around the systemic velocities (6.4 -- 6.7 km s$^{-1}$), the model channel maps show that the CS emission is extended
on both the northern and southern sides with eastern and western emission peaks.
The observed velocity channel maps in these velocities also show that the CS emission is extended
in both the northern and southern sides, although the eastern emission peaks seen in the model channel maps
are not clear in the observed velocity channel maps.
These results suggest that the observed velocity structures in the lower-velocity region
can be interpreted as a rotating and infalling envelope,
which has a rotational profile with the conserved angular momentum and
the infalling speed slower
than that expected from the free-fall onto the central stellar mass of 0.5 $M_{\odot}$.

By contrast, the high-velocity CS emission is unlikely to trace the same envelope kinematic component.
We performed $\chi^2$ model fitting of the rotating and infalling envelope
to the high-velocity CS emission, which is
expressed by eq. (13), with $M_\star^{inf}$, $j$ and $\theta$ as fitting parameters,
and fixed $v_{sys}=6.6$ km s$^{-1}$, $i=60\degr$, and $\sigma_{gas}=0.7$ km s$^{-1}$.
The best-fit result of the rotating and infalling envelope model to the high-velocity CS emission
is $M_\star^{inf}$ = 8.0 $\times$ 10$^{-6}$ $M_\odot$, $j$ = 250 AU km s$^{-1}$, and $\theta=-40\degr$,
which corresponds to $v_{rad}$ = 0.01 km s$^{-1}$ and $v_{rot}$ = 1.8 km s$^{-1}$ at $r=$140 AU.
This result is unphysical because this mass is too small for a typical Class I protostar,
and previous studies on the orbital motion of L1551 IRS 5 with VLA show the total mass $\sim$0.9 $M_\odot$ \cite{lim06}.
In fact this fitting result demonstrates that there is little infall, which means the motion is essentially Keplerian,
and thus our interpretation of the Keplerian disk is more straightforward to reproduce the observed
velocity feature of the high-velocity CS emission.

In summary, the submillimeter CS emission in L1551 IRS 5 traces primarily two distinct
components; one is an $r\sim$64 AU-scale Keplerian circumbinary disk,
and the other an extended ($r\sim$500 AU) rotating and infalling envelope
with the $r^{-1}$ rotational profile and reduced infalling speed.
In the next section, we will discuss these results in the context of disk formation
from protostellar envelopes and evolution of disks around protostellar sources.

\section{Discussion}
\subsection{Keplerian Disk Embedded in the Infalling Envelope around L1551 IRS 5}

As described in the previous section, our SMA + ASTE observations of the Class I
protostellar binary L1551 IRS 5 in the 343 GHz continuum and CS (7--6) line
have revealed an $r\sim$64 AU-scale Keplerian disk with a mass of $\sim0.070$ $M_{\odot}$
around the central protobinary mass of $\sim$0.5 $M_\odot$,
which is embedded in a rotating and infalling envelope.
The observed rotational profile in the infalling envelope is consistent with rotation with
the conserved angular momentum found in the outer ($r\sim$1200 AU) envelope
($i.e.$, eq. (9); Momose et al. 1998), and connects to the inner Keplerian rotation.
On the other hand,
the observed infalling velocity is clearly lower than the free-fall velocity onto
the central protobinary mass of 0.5 $M_\odot$, which is derived from the observed
Keplerian rotation in the disk. We consider that
these results reveal the transition from the infalling envelope to the Keplerian disk.

\begin{sloppypar}
Although it is not straightforward to form Keplerian disks
from infalling envelopes
because of the efficient magnetic braking \cite{mel08,mel09,ma11a,ma11b,zli11}, 
the latest theoretical simulations \cite{mac14} successfully
produced Keplerian disks with properties similar to those found by the previous and present observations.
They argued that the outcome of their simulations highly depends on the initial model settings such as the sink accretion radius and
the initial core density structure (uniform or Bonnor-Ebert sphere).
For their small sink accretion radii
($r_{acc} \lesssim$1 AU), they produced $r\sim$50--100 AU Keplerian disks
with masses of $\sim$0.06--0.15 $M_{\odot}$
embedded within $r\sim$1000-AU scale flattened envelopes,
when the central protostellar masses reach $\sim$0.5 $M_{\odot}$.
In the outer region ($r$ $\gtrsim$200 AU) the envelope rotation approximately
follows the $r^{-1}$ profile that connects to the inner
Keplerian rotation.
In the intermediate region (50--100 AU $\leq$ $r$ $\leq$ 200 AU), the decrease in the infalling speed
is also reproduced, as the matter approaches the centrifugal radius.
At the stage of the formation of $r$ $\sim$50--100 AU Keplerian disks,
substantial envelope masses ($\sim$1 $M_{\odot}$) still remain.
Furthermore, the morphology of the infalling envelopes at $r$ $\sim$1000 AU
scale is flattened with the typical scale height of 100--200 AU \cite{mac14}.
These results show that our observational results in L1551 IRS 5,
as well as those of the C$^{18}$O (1--0) emission \cite{mom98},
can be reproduced by the latest theoretical simulations.
\end{sloppypar}

The picture of the transition from the infalling envelope to the Keplerian disk shows certain similarities among different sources.
In the case of L1551 NE, another Class I protobinary source located
$\sim$2$\farcm$5 northwest of L1551 IRS 5,
Takakuwa et al. (2013) have found that the infalling velocity is much smaller
than the free-fall velocity toward the central stellar mass of 0.8 $M_\odot$ derived from the Keplerian rotation.
The reduction of the infalling velocities is also found in a Class 0/I protostar L1527 IRS \cite{oha14}
and a Class I protostar TMC-1A (Aso et al. in prep) from ALMA Cycle 0 observations.
On the other hand, the observed lower angular momentum of the infalling gas than that of the central disk in L1551 NE
is essentially different from the result of L1551 IRS 5.
As discussed by Takakuwa et al. (2013), in the case of L1551 NE the magnetic braking
may be efficient in the envelope region but inefficient in the disk due to the higher
plasma $\beta$, the ratio of the gas thermal pressure to the magnetic pressure.
Around a Class I protostar L1489 IRS, Yen et al. (2014) have identified free-falling gas with the rotational
angular momentum consistent with that of the outer boundary of the Keplerian disk.
In the case of L1489 IRS the amount of the surrounding envelope is much smaller ($\sim$0.02 $M_{\odot}$)
than that of L1551 IRS 5 and L1551 NE, and thus the free-falling gas component
may be a remnant gas component falling toward the central disk.
Accumulation of more examples of the transitions from infall to Keplerian rotation
is needed to draw a more general picture of disk formation from protostellar envelopes.
Furthermore,
the transitional regions from the infalling envelopes to the Keplerian disks
are expected to be at $r$ $\sim$100--300 AU.
To fully resolve such size scales,
higher-resolution observations at $\sim$10 AU are needed.
Forthcoming ALMA observations of protostellar envelopes should be able to achieve this goal.

\subsection{Evolution of Keplerian Disks around Protostars}

Our observations of L1551 IRS 5 unveiled a Keplerian disk around the protobinary
embedded in the infalling envelope. Since the mass of the entire envelope around L1551 IRS 5
derived from the single-dish 850 $\micron$ mapping is $\sim$1 $M_{\odot}$ \cite{mor06}
and thus larger than the disk ($\sim$0.070 $M_{\odot}$) + protobinary ($\sim$0.5 $M_{\odot}$) masses,
further growth of the central protobinary + disk system through the mass accretion from the envelope
is expected.
On the assumption that the high-velocity CS emission (Figure \ref{brcs} \emph{left})
traces the same disk component as that traced by the 343 GHz continuum emission,
and that the CS emission is in the LTE condition with the excitation temperature of
47 K \cite{mor94},
we derived the CS abundance to be $\sim$1.6 $\times$ 10$^{-8}$.
Applying this abundance to the low-velocity CS emission (Figure \ref{brcs} \emph{right})
yields the envelope mass of $\sim$0.38 $M_\odot$. The typical mean radius of the envelope
component is $\sim$3$\farcs$5, and the infalling velocity at that radius is $\sim$0.43 km s$^{-1}$,
providing the infalling time of $\sim$5400 $yr$ and thus the mass accretion rate of
$\sim$7.0 $\times$ 10$^{-5}$ $M_\odot$ $yr^{-1}$.
On the other hand, the mass outflowing rate has been estimated to be
$\sim$1 $\times$ 10$^{-6}$ $M_\odot$ $yr^{-1}$ by Liseau et al. (2005).
Therefore, the envelope material can further accrete onto the central
system during the typical Class I period of 2.5 -- 6.7 $\times$ 10$^{5}$ yr \cite{hat07}.

%
%
%

Recent high-resolution interferometric observations
have been finding more and more Keplerian circumstellar disks around
protostellar sources embedded in protostellar envelopes
\cite{bri07,lom08,jor09,tob12,yen13,har14}.
Our previous SMA observations of L1551 NE have identified an
$r$ $\sim$300-AU scale Keplerian disk with a central stellar mass of 0.8 $M_{\odot}$ \cite{tak12,tak13},
embedded in an envelope with the mass of $\sim$0.4 $M_{\odot}$ \cite{mor06}.
With ALMA, more deeply embedded Keplerian disks are found around Class 0 protostars such as
VLA 1623A, HH 212, and L1527 IRS \cite{mur13,lee14,oha14}.
Yen et al. (2013) investigated the evolutionary sequence of protostellar envelopes
into Keplerian-disk formation, based upon their SMA survey of protostellar envelopes,
as well as previously published results.
They found that gas motions in the protostellar envelopes can be categorized into three types;
(1) no detectable rotation but infalling motion around early Class 0 protostars;
(2) $v_{rot}$ $\sim$$r^{-1}$ rotational motion in the infalling envelopes around late
Class 0 and early Class I sources; and (3) large-scale Keplerian disks around Class I
protostars. They proposed that the differences of the envelope gas motions reflect
evolutionary trends of the protostellar envelopes into the Keplerian-disk formation.
L1551 IRS 5 is classified in category (2) in their paper, but our new higher-resolution
observations unveiled the Keplerian disk embedded in the infalling envelope.
These results imply that Keplerian circumstellar disks with size scales comparable to
those around T-Tauri stars are present
at the protostellar stages still deeply embedded in protostellar envelopes, and
that Keplerian-disk formation should occur in the late Class 0 or early Class I stages.

To discuss the evolutionary sequence of Keplerian disks around protostellar sources
after their formation, in Table \ref{tbl-3} we compile the measured physical
properties of the protostars and the associated Keplerian disks, including those of
L1551 IRS 5, sorted by the bolometric temperature ($T_{bol}$).
As another evolutionary indicator, we also list the ratio of the central protostellar mass
to the total star+disk+envelope mass
($\equiv$ $\frac{M_{\star}}{M_{\star}+M_{disk}+M_{env}}$)
in Table \ref{tbl-3}.
The mass ratio of L1551 IRS 5 is the lowest among the Class I sources,
and that of the two Class 0 sources are the lowest among the entire sample
as expected. However, there is no clear distinction of the radii and masses of the Keplerian disks
between these early protostellar sources and the other, more evolved protostars.
%
These results suggest that the Keplerian disks do not grow much
after the disks become hundreds AU scales at the late Class 0 or early Class I stages.
In contrast, the positive trend of the ratio of the central protostellar mass
to the total star+disk+envelope mass with $T_{bol}$ implies
further growth of the central protostellar masses with respect to the envelope masses.
The Keplerian disks may be in a quasi-static mode of mass accretion onto the protostars,
or episodic mass accretions from the disks onto the central protostars may be ongoing,
as proposed from Spitzer survey observations of protostellar sources \cite{dun08}.
Theoretical simulations by Machida \& Hosokawa (2013) and Machida et al. (2014)
show that the mass of the Keplerian disk does not grow much and saturates at
$\sim$0.06--0.15 $M_{\odot}$ after the large-scale $r \sim$50--100 AU Keplerian disk appears,
while the mass of the central protostar keeps growing from $\sim$0.5 $M_{\odot}$ to
$\sim$1.0 $M_{\odot}$. In the innermost part of the disk ($r \lesssim$30 AU)
mass accretion toward the central protostar is reproduced because of the slow
angular momentum transfer by the outflows.
Thus, these theoretical simulations are at least qualitatively consistent with the observed
evolutionary properties of the Keplerian disks around the protostellar sources.

The Keplerian disk around L1551 IRS 5 is a circumbinary disk surrounding a
protostellar binary system. Since the radius of the Keplerian circumbinary disk is $\sim$64 AU
and the projected binary separation $\sim$40 AU, dynamical interaction between
the circumbinary disk and the protobinary is anticipated.
In the more evolved (T-Tauri) binary GG Tau (projected separation $\sim$35 AU),
a ringlike feature of the circumbinary disk,
where the innermost part of the disk ($r$ $\lesssim$200 AU) has been cleared by
the dynamical interaction with the binary stars, is identified \cite{gui99,and14}.
Recently, such a ringlike feature is also
identified in a protostellar binary L1551 NE with ALMA by Takakuwa et al. (2014).
The observed ringlike structure and the gas motion in the circumbinary disk of L1551 NE
can be reproduced with spiral-arm features and 
rotating and accreting gas motions toward the protostellar binary,
caused by the gravitational torques from the binary protostars.
The spatial resolution of the present SMA observations of L1551 IRS 5
($\sim$100 AU) is, however, not high enough to separate the circumbinary ring
from the circumstellar disks around the individual stars and investigate whether
the dynamical interaction between the circumbinary disk and the binary is present
or not. Further higher-resolution observations of L1551 IRS 5 will unveil the
presence/absence of such a dynamical interaction in this protostellar binary system.

\section{Summary}

We have carried out sub-arcsecond resolution SMA observations of the protobinary system L1551 IRS 5
in the 343 GHz continuum and the CS (7--6) line. The SMA data in the CS (7--6) line
are combined with the ASTE single-dish data to fill in the missing short spacing information,
and the combined image cube is discussed with our simple kinematical models.
The main results are summarized below:

1. The 343 GHz dust-continuum emission in L1551 IRS 5 shows an intense central peak
associated with the protobinary with an elongated feature along the northwest to southeast direction
(160$\pm$10 AU $\times$ 80$\pm$10 AU, P.A. = -23$\degr\pm$10$\degr$),
which is perpendicular to the associated binary jets.
The continuum emission most likely traces the circumbinary material, and
the mass of the circumbinary material is estimated to be $\sim$0.070 $M_{\odot}$.

2. The combined SMA + ASTE moment 0 map of the CS emission in L1551 IRS 5
shows a $\sim$1000 AU-scale feature with its peak close to the protobinary.
The CS velocity channel maps reveal that the velocity structures can be decomposed
into two components. One is a high-velocity ($\gtrsim$1.5 km s$^{-1}$),
compact blueshifted and redshifted emission located
to the southeast and northwest of the protobinary, respectively, which traces the structure
of the circumbinary material seen in the 343 GHz dust-continuum emission and
exhibits the velocity gradient along the major axis of the continuum emission.
The other is a low-velocity ($\lesssim$1.3 km s$^{-1}$), extended blueshifted and redshifted emission,
that exhibits a slight south (blueshifted) to north (redshifted) emission offset, but shows
emission overlaps.

3. $\chi^2$ model fitting of geometrically-thin Keplerian disks to
the 3-dimensional image cube of the high-velocity CS emission was performed.
The channel maps of the high-velocity CS emission are well-reproduced by a
geometrically-thin Keplerian disk model with the central stellar mass
$M_\star$ = 0.5$^{+0.6}_{-0.2}$ $M_\odot$ and the
disk position angle $\theta$ = -33$\degr$$^{+17\degr}_{-24\degr}$, where
the disk inclination angle $i$ = -60$\degr$ and the internal velocity dispersion $\sigma$ = 0.7 km s$^{-1}$
are fixed.
The extent and the major axis of the best-fit Keplerian disk model match well
with those of the 343 GHz continuum emission. These results show that
both the high-velocity CS emission and the 343 GHz dust-continuum emission trace
a Keplerian disk around the protobinary.

4. The observed velocity features of the low-velocity CS emission component
can be reproduced with a model of the flattened rotating and infalling envelope.
The rotational profile is consistent with the rotation of the conserved angular momentum
deduced from the previous C$^{18}$O (1--0) observations ($i.e.$, $v_{rot}$ = $\frac{j}{r}$;
$j$ = 168 AU km s$^{-1}$), and connects smoothly to the inner Keplerian rotation at
a radius of $\sim$64 AU, which is consistent with the radius of the 343 GHz continuum emission.
The derived infalling velocity is 3 times smaller than the free-fall velocity (10 times smaller in terms of
the central stellar mass)
toward the central stellar mass of 0.5 $M_{\odot}$ as estimated from the Keplerian rotation,
suggesting that the infall velocity is reduced as the matter approaches close to the centrifugal radius.
We suggest that these results demonstrate the transition from the outer
infalling envelope to the inner Keplerian disk.

5. The latest theoretical models of disk formation from protostellar envelopes can reproduce
the main characteristic features of the observed transition from the infalling envelope to the inner
Keplerian disk in L1551 IRS 5, including the disk radius ($\sim$64 AU), mass of the central object ($\sim$0.5 $M_{\odot}$),
disk ($\sim$0.070 $M_{\odot}$), and the envelope ($\sim$1 $M_{\odot}$),
and the $v_{rot}$ $\sim$$r^{-1}$ rotational profile and the reduced infalling speed in the envelope.
Compilation of protostellar sources associated with Keplerian disks shows that the Keplerian
disks do not grow much once they become large enough to be observable ($r$ $\gtrsim$50--200 AU)
at the late Class 0 or the early Class I stages,
while the central protostars appear to keep growing through mass accretion.
These results indicate that the Keplerian disks around the protostars keep supplying materials
to the central objects, whereas the disks themselves do not grow. These results are also consistent
with the latest theoretical models of disk formation, which show saturation of the masses of the Keplerian disks
at around $\sim$0.06--0.15 $M_{\odot}$ after the disks grow to large scales ($r \sim$50--100 AU),
while the masses of the central protostars keep growing from $\sim$0.5 $M_{\odot}$ to
$\sim$1.0 $M_{\odot}$ through the mass supply from the innermost part of the disks.

\acknowledgments
We are grateful to J. Lim, M. Momose, M. Saito, and Zhi-Yun Li for their fruitful discussions,
and an anonymous referee for helpful suggestions.
We would like to thank all the SMA staff supporting this work.
S.T. acknowledges a grant from the National Science
Council of Taiwan (MOST 102-2119-M-001-012-MY3) in support of this work.

\clearpage

\clearpage
\begin{deluxetable}{lcc}
\tabletypesize{\scriptsize}
\tablecaption{Parameters for the SMA Observations of L1551 IRS 5 \label{tbl-1}}
\tablewidth{0pt}
\tablehead{\colhead{Parameter} & \multicolumn{2}{c}{Value}\\
\cline{2-3}
\colhead{} & \colhead{SMA extended} & \colhead{SMA subcompact} }
\startdata
Observation Date & 2012 Dec 23, 2013 Jan 4 & 2009 Sep 21 \\
Number of Antennas &7 &7 \\
Field Center&
  \multicolumn{2}{c}{(04$^{\rm h}$31$^{\rm m}$34$\fs$14, 18$^{\circ}$08$\arcmin$05\farcs1)} \\
Primary Beam HPBW& \multicolumn{2}{c}{$\sim$36$\arcsec$}\\
Synthesized Beam HPBW (Continuum, SMA)&
  \multicolumn{2}{c}{0\farcs77 $\times$ 0\farcs53 (P.A. = -88$\degr$)}\\
Synthesized Beam HPBW (CS line, SMA+ASTE)&
  \multicolumn{2}{c}{0\farcs90 $\times$ 0\farcs65 (P.A. = 82$\degr$)}\\
Baseline Coverage &{20.7 -- 218.0 (m)} &{6.7 -- 67.3 (m)}\\
Frequency Coverage & 328.9 -- 332.9 GHz (LSB) & 342.8 -- 344.8 GHz (LSB)\\
 & 340.9 -- 344.9 GHz (USB) & 352.8 -- 354.8 GHz (USB)\\
Conversion Factor (CS line, SMA+ASTE) & \multicolumn{2}{c}{1 (Jy beam$^{-1}$) = 17.8 (K)}\\
Frequency Resolution & \multicolumn{2}{c}{203 kHz $\sim$0.178 km s$^{-1}$}\\
Flux Calibrator &Neptune &Mars\\
Gain Calibrator &{0423-013, 0530+135} &{0423-013, 3c120}\\
Flux (0423-013) &1.82 Jy &3.65 Jy\\
Flux (0530+135)       &0.30 Jy & \\
Flux (3c120)       &  &0.86 Jy\\
System Temperature (DSB) &$\sim$350 -- 1250 K &$\sim$180 -- 330 K\\
rms noise level (Continuum, SMA)& \multicolumn{2}{c}{8.4 mJy beam$^{-1}$}\\
rms noise level (CS Line, SMA+ASTE)& \multicolumn{2}{c}{0.130 Jy beam$^{-1}$}\\
\enddata
\end{deluxetable}

\clearpage
\begin{deluxetable}{lcccccccc} \tablecaption{Prototellar Sources with Keplerian Disks \label{tbl-3}}
\setlength{\tabcolsep}{4pt}
\tablewidth{0pt}
\tablehead{\colhead{Source} &\colhead{$L_{bol}$} &\colhead{$T_{bol}$} &\colhead{$R_{kep}$$^a$} &\colhead{$M_{\star}$$^b$} &\colhead{$M_{disk}$$^c$} &\colhead{$M_{env}$$^d$} &\colhead{$\frac{M_{\star}}{M_{\star}+M_{disk}+M_{env}}$} &\colhead{References$^e$}\\
           \colhead{} &\colhead{($L_\odot$)} &\colhead{($K$)} &\colhead{(AU)} &\colhead{($M_\odot$)} &\colhead{($M_\odot$)} &\colhead{($M_\odot$)} &\colhead{($\%$)} &\colhead{}}
\startdata
\multicolumn{9}{c}{Class 0 sources} \\ \hline
VLA1623A    &  1.1  &  10 & 180  & 0.22    & ...             & 0.8       & $<$22     & 1,2  \\
L1527 IRS   &  1.97 &  44 &50--90&0.19--0.3& 0.007           & 0.9       & 17        & 3,4,5 \\ \hline
\multicolumn{9}{c}{Class I sources} \\ \hline
L1551 NE    &  4.2  &  91 & 300  & 0.8     & 0.026           & 0.39      & 65        & 2,6,7 \\
L1551 IRS 5 & 22.1  &  94 &  64  & 0.5     & 0.070           & 1.01      & 32        & This work,3,6 \\
TMC1        &  0.9  & 101 & 100  & 0.54    & 0.005--0.024    & 0.14      & 78        & 3,8     \\
TMC1A       &  2.7  & 118 & 100  & 1.1     & 0.025--0.06     & 0.12      & 88        & 3,8,9 \\
L1489-IRS   &  3.7  & 238 & 200  & 1.8     & 0.007--0.02     & 0.11      & 95        & 9,10,11   \\
L1536       &  0.4  & 270 &  80  & 0.4     & 0.007--0.024    & 0.05      & 86        & 3,8,12  \\
Elias 29    & 14.1  & 299 & 200  & 2.5     & $\lesssim$0.007 & 0.025     &$\gtrsim$99& 3,13 \\
IRS 43      &  6.0  & 310 & 140  & 1.0     & 0.0081          & 0.026     & 97        & 11     \\
IRS 63      &  1.0  & 327 & 100  & 0.37    & 0.055           & 0.022     & 83        & 3,13  \\
\enddata
\bigskip
\begin{flushleft}
\textbf{Notes.} \\
$^a$Outer radius of the Keplerian disk. \\
$^b$Central protostellar mass. \\
$^c$Mass of the Keplerian disk. \\
$^d$Mass of the protostellar envelope within the radii of $\sim$10$\arcsec$--15$\arcsec$.\\
$^e$References: (1) Murillo et al. 2013; (2) Froebrich 2005; (3) Kristensen et al. 2012; \\
(4) Tobin et al. 2012; (5) Ohashi et al. 2014; (6) Moriarty-Schieven et al. 2006; \\
(7) Takakuwa et al. 2012; (8) Harsono et al. 2014; (9) Yen et al. 2013; (10) Brinch et al. 2007; \\
(11) J\o rgensen et al. 2009; (12) Eisner 2012; (13) Lommen et al. 2008.
\end{flushleft}
\end{deluxetable}

\end{document}